\documentclass[a4paper,11pt]{article}
\pdfoutput=1 % if your are submitting a pdflatex (i.e. if you have
\usepackage{jcappub} % for details on the use of the package, please
\usepackage{amsmath}
\usepackage{amssymb} 
\usepackage{graphicx}
\usepackage{enumerate}
\usepackage{color}
\usepackage{tensor}
\usepackage{mathrsfs}
\usepackage{hyperref}
\usepackage[normalem]{ulem}
\usepackage[dvipsnames]{xcolor}
\usepackage{accents}
\usepackage{scalerel}
\usepackage{bm}
\usepackage{comment}
\DeclareMathOperator{\diag}{diag}
\newcommand{\lambdabar}{{\mkern0.75mu\mathchar '26\mkern -9.75mu\lambda}}

\makeatletter
\newcommand*{\leqdef}{\mathrel{\rlap{%
			\raisebox{0.25ex}{$\m@th\cdot$}}%
		\raisebox{-0.25ex}{$\m@th\cdot$}}%
	=}
\newcommand*{\reqdef}{=\mathrel{\rlap{%
			\raisebox{0.25ex}{$\m@th\cdot$}}%
		\raisebox{-0.25ex}{$\m@th\cdot$}}
}
\makeatother

\title{Influence of gravitational waves upon light. Part II. Electric field propagation and interference pattern in a gravitational wave detector}
\author[a]{Jo\~ao C. Lobato,}

\author[a]{Isabela S. Matos,}

\author[a]{Lucas T. Santana,}

\author[a, b]{Ribamar R. R. Reis}

\author[a]{and Maur\'\i cio O. Calv\~ao}

\affiliation[a]{Universidade Federal do Rio de Janeiro,
	Instituto de F\'\i sica, \\
	CEP 21941-972 Rio de Janeiro, RJ, Brazil}

\affiliation[b]{Universidade Federal do Rio de Janeiro, Observat\'orio do Valongo, 
	\\CEP 20080-090 Rio de Janeiro, RJ, Brazil}

\emailAdd{jcavlobato@if.ufrj.br}
\emailAdd{isa@if.ufrj.br}
\emailAdd{lts@if.ufrj.br}
\emailAdd{ribamar@if.ufrj.br}
\emailAdd{orca@if.ufrj.br}

\abstract{In this second article of the series, we apply our recently derived equation for the electric field propagation along light rays \cite{Santana2020}, valid on the electromagnetic geometrical optics limit, to the special case of a toy interferometer used to detect gravitational waves in a flat background. Such an equation shows that, assuming the detector is in the transverse-traceless frame, which has a local shearing relative motion due to the gravitational wave perturbations, the electric field does not propagate as in an inertial reference frame in Minkowski spacetime. We present the electric field at the end of the interferometric process, for arbitrary arm configurations with respect to the plane gravitational wave packet propagation direction. Then, for normal incidence, we compute the interference pattern and, in addition to the usual term associated with the difference in path traveled by light in the arms, we deduce two new contributions to the final intensity, arising from: (i) the round-trip electromagnetic frequency shift and (ii) the divergence of the light beam. Their quantitative relevance is compared to the traditional contribution and shown to be typically negligible due to the geometrical optics regime of light. Moreover, a non-parallel transport of the polarization vector takes place, in general, because of the gravitational wave, a feature which could generate further contributions. However, we conclude that for the normal incidence case such vector is parallel transported, preventing this kind of correction.}


\begin{document}

\maketitle
\flushbottom
\section{Introduction}
\label{sec:intro}

In the first paper of this series \cite{Lobato2021Part1} (hereafter L1), we presented two quantities relevant to this second work and discussed how they change when the influence of a linearized gravitational wave (GW) on Minkowski background upon an electromagnetic (EM) wave is investigated: (i) the radar distance between two transverse-traceless (TT) observers, concluding that it is not affected by light spatial trajectory perturbations due to the GW; (ii) light's frequency shift acquired after a round-trip between TT observers. Despite the existence of this Doppler effect, we argued that, assuming that the interference pattern of a GW detector depends only on the difference in phase of the interacting beams at the end of the interferometric process, no extra contribution to the final intensity arises from that frequency shift, as one might suspect from heuristic arguments \cite{Saulson1997,Faraoni2007}. In this second part, this assumption will be the exact subject of scrutiny, once we allow the electric field propagation to include the full curved nature of spacetime and the kinematics of the TT frame, enumerated in L1 as facet (iv) of the action of GWs on EM waves.

Following the usual Michelson-Morley interferometric experiment, the intensity pattern in a GW detector is commonly computed \cite{Maggiore2007} in terms of the phase difference between the two beams at the recombination event $\mathcal{D}$ (cf. the right panel of figure \ref{fig:toymodel}), being directly related to the difference in radar distances of the two arms (cf. subsection V.C of L1). To that end, together with the constancy of phase along light rays, one of the EM geometrical optics laws, the simple propagation equation, even in TT coordinates, is assumed
\begin{equation}
	\frac{dE^{\mu}}{d\vartheta} = 0\,, \label{eq: Mink_Elect_evol}
\end{equation}
where $\vartheta$ is the affine parameter of the null geodesic of choice. In other words, the magnitude and polarization of the electric field $E^{\mu}$ are bound to evolve freely in each arm of the interferometer, as in an inertial frame of Minkowski spacetime, even though GWs are passing by. This would justify disregarding the connection coefficients that bring eq.~(\ref{eq: Mink_Elect_evol}) to covariant form, since in Minkowski spacetime there is always a coordinate system with vanishing connection to which a given Minkowski inertial frame is adapted.

In our recent paper \cite{Santana2020}, as an effort to clear out the laws determining light propagation with respect to a definite arbitrary set of observers in a generic spacetime, we have shown that the electric field of an EM wave in the geometrical optics approximation of Maxwell equations (cf. appendix \ref{app:geometrical_optics}) evolves along any of its light rays according to
\begin{equation}
	\frac{DE^{\mu}}{d\vartheta} + \frac{1}{2} \widehat{\Theta} E^{\mu} = \left( \frac{k^{\mu} E^{\nu} - k^{\nu} E^{\mu}}{\omega_{\textrm{e}}}\right) \frac{Du_{\nu}}{d\vartheta}\,,	
	\label{eq:electric_evolution}
\end{equation}
where $\omega_{\textrm{e}}\leqdef - k^{\mu} u_{\mu}$ is the frequency of light, $u^{\mu}$ is the 4-velocity of the frame, $k^{\mu}$ is the null tangent vector of the rays, $\hat{\Theta}$ is the optical expansion of the beam and, of course, for any vector $v^{\mu}$ along the ray, $Dv^{\mu}/d \vartheta \leqdef dv^{\mu}/d \vartheta + \Gamma^{\mu}_{\alpha \gamma} v^{\alpha} k^{\gamma}$.

In contrast to eq.~(\ref{eq: Mink_Elect_evol}), eq.~(\ref{eq:electric_evolution}) shows three possibly relevant contributions to the electric field evolution: (i) the connection coefficients on the left-hand side (LHS); (ii) the optical expansion and (iii) the frame kinematics on the right-hand side (RHS). It becomes expedient, then, to evaluate what is the role played by each of them on the calculation of the interference pattern of a GW detector. 

To illustrate this, we here explicitly compute and elaborate more on the application subject appearing in \cite{Santana2020}: the electric field at the end of the round-trips in each of the two beams on a toy GW Michelson-Morley-like interferometer adapted to the TT coordinates, whose anisotropic stretch and squeeze give rise to terms (i) and (iii) (cf. section \ref{sec:kinematics}). The radar distance and the frequency shift will appear in the final intensity pattern, which will give us the opportunity to bridge our result with the ones shown in L1, and hence to provide clarifying insights on the physical origins surrounding the interference pattern fluctuations in the current context.

We find that, for GW normal incidence, there are other contributions to the intensity pattern rather then the usual one related to the difference in path traveled by light in both arms of the interferometer (also related with the phase difference of the beams) and compare the relevance of each of these terms using some LIGO-like experimental parameters. Furthermore, we are able to determine whether the non-parallel transport of the polarization vector (the direction of the electric field as considered in \cite{Santana2020}) and the non-trivial propagation of the  electric field amplitude have, separately, a consequence on the interference pattern.

In section \ref{sec:description}, we describe the interferometry procedure we aim to analyze. In section \ref{sec:kinematics}, the TT frame kinematics is examined. The optical expansion is propagated along the rays and its initial value is estimated assuming aLIGO parameters. In section \ref{sec:electriceq}, the terms of eq.~(\ref{eq:electric_evolution}) are physically discussed. Section \ref{sec:electricsolutions} deals with the solutions for such equation, under the mixed conditions presented in L1, for a beam traveling one of the arms of the interferometer that is assumed to be in an arbitrary orientation with respect to the GW propagation direction. Finally, in section \ref{sec:finalintensity} we compute the final intensity measured once the two beams interfere for the case of a normally incident GW.

%\begin{table*}
%	\caption{\sc{Symbols used in this series of articles.}}
%	\label{tab:symbols}	
%	\begin{tabular}{|c|l|c|}	
%		%\begin{longtable*}{|c|l|c|}
%		\hline
%		\textbf{Symbol} & \hspace{3.5cm} \bfseries{Description} & \textbf{Reference(s)} \\
%		\hline 
%		$a$ & Acceleration of $u$ & \eqref{irrdecomp} \\
%		$\Theta$ & Expansion scalar of $u$ & \eqref{irrdecomp}  \\
%		$\sigma$ & Shear tensor of $u$ & \eqref{irrdecomp}  \\
%		$\Omega$ & Vorticity tensor of $u$ & \eqref{irrdecomp}  \\
%		$p$ & Projector tensor (onto the rest space of $u$) & \eqref{projector} \\
%		$s$ & Screen space projector (associated to $u$ and $k_{(\bm{\epsilon})}$) & \eqref{screen_projector} \\
%		$\hat\Theta$ & Optical expansion scalar of a null geodesic field & ??? \\
%		$\hat{\sigma}$ & Optical shear tensor of a null geodesic field &  ??? \\
%		$\hat{\Omega}$ & Optical vorticity tensor of a null geodesic field &  ??? \\
%		$E$ & Electric vector field or its scalar magnitude & \eqref{electric_field} \\
%		$e$ & Unit vector along electric field & ??? \\
%		$F$ & Faraday (electromagnetic) tensor & \eqref{Maxwell} \\
%		$\psi$ & Electromagnetic phase in eikonal approximation & \eqref{ansatz} \\
%		$f$ & Electromagnetic amplitudes in eikonal approximation & \eqref{ansatz} \\
%		$E_T$ & Total (superposed) electric field & ??? \\
%		$I_T$ & Total (superposed) electric intensity & ??? \\
%		\hline	
%		%\end{longtable*}
%	\end{tabular}
%\end{table*}

On this second part of our study, we will continue to symbolize the TT coordinates by the 4-tuple $(t,x,y,z)$.  The TT metric is given by:
\begin{align}
g_{\alpha \beta}(t - x) &\leqdef  \eta_{\alpha \beta} + \epsilon_P h^P_{\alpha \beta}(t - x)\,, \label{eq:metric} \\
\eta_{\alpha\beta} &\leqdef \diag(-1, 1,1,1)\,,
\end{align}
where the two GW polarizations are indexed by $P= +, \times$, for which the usual Einstein summation convention holds, and
\begin{align}
h^+_{\alpha \beta}(t-x) & \leqdef h_{+}(t-x)(\delta_{\alpha z}\delta_{\beta z} - \delta_{\alpha y}\delta_{\beta y})\,, \label{eq:hplus} \\
h^{\times}_{\alpha \beta}(t-x) & \leqdef -h_{\times}(t-x) (\delta_{\alpha y} \delta_{\beta z}+\delta_{\alpha z} \delta_{\beta y})\,. \label{eq:hcross}
\end{align}
All calculations will be performed up to linear order in $\bm{\epsilon}\leqdef (\epsilon_+, \epsilon_{\times})$ and the notation used in L1 will be adopted here as well, including the subscript ($\bm{0}$) indicating zeroth order terms in an $\bm{\epsilon}$ expansion. 
\section{Interferometry description}
\label{sec:description}

\begin{figure}[t]
	\centering
	\includegraphics[width=0.38\textwidth]{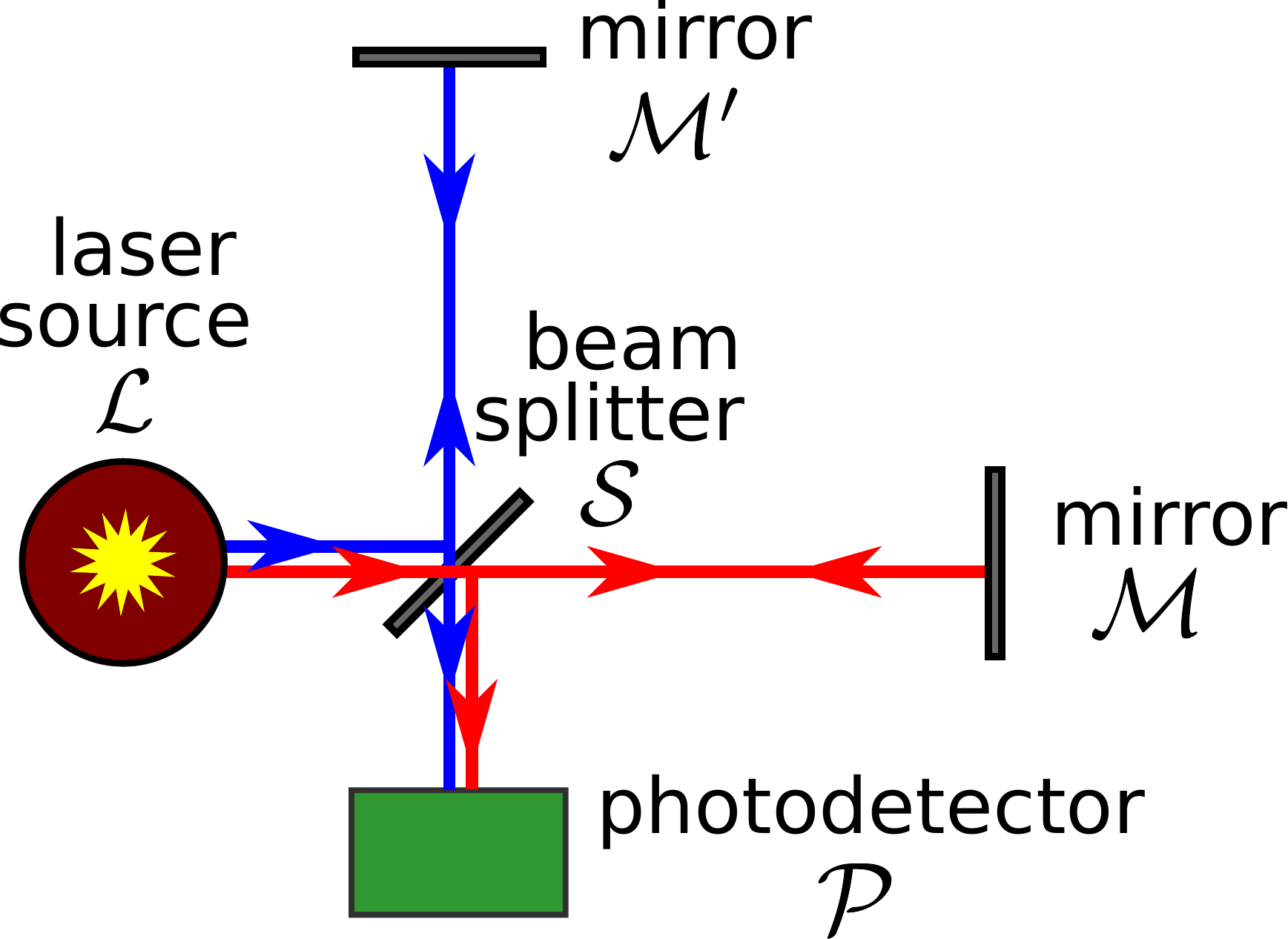} \hspace{25pt}
	\includegraphics[width=0.38\textwidth]{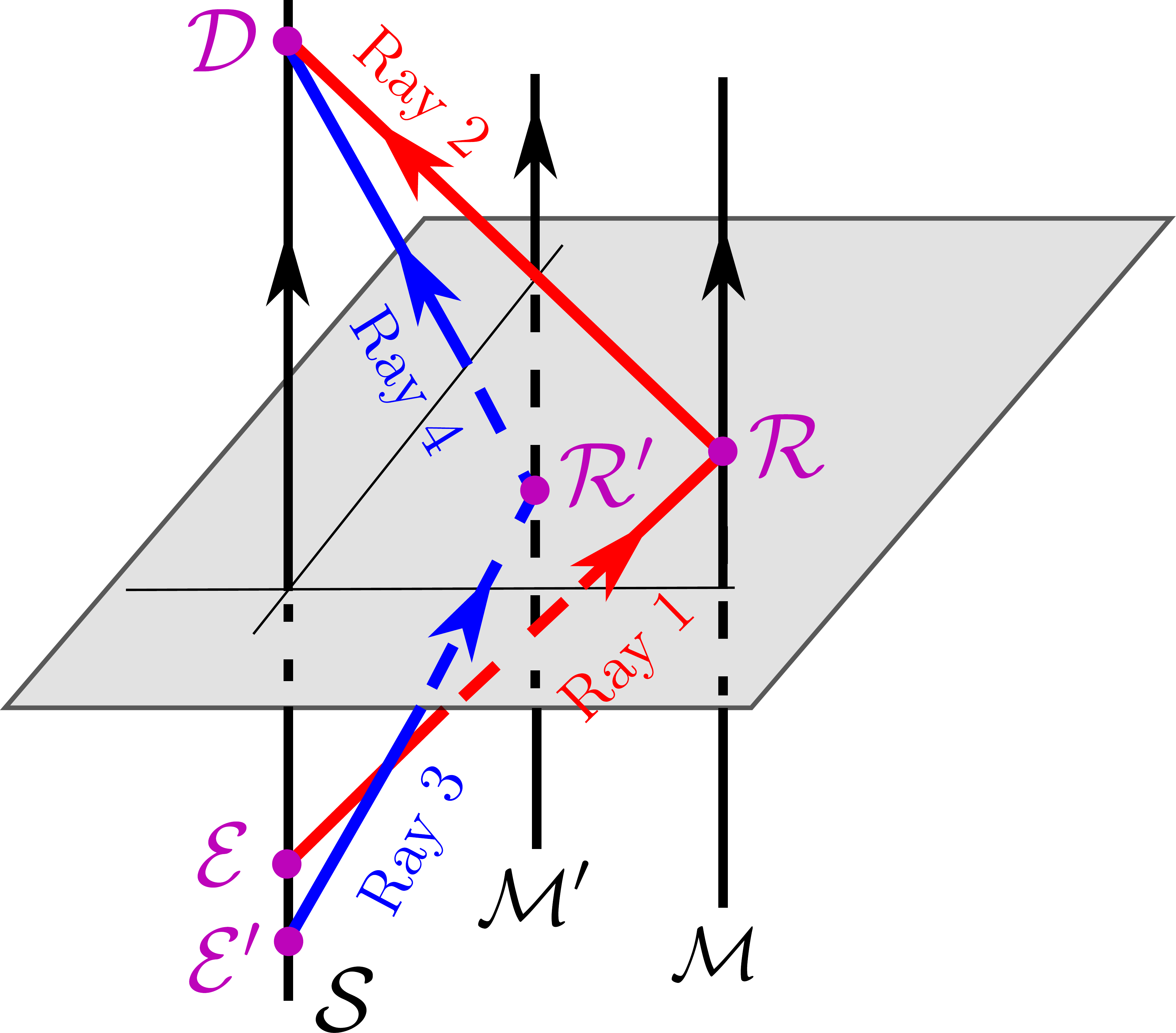}
	\caption{\emph{Left}: Spatial diagram of a Michelson-Morley GW interferometer. \emph{Right}: Corresponding spacetime diagram, with relevant worldlines: tilted ones for the null geodesic arcs and vertical ones, for the several devices. Although the above images illustrates perpendicular arms, we here assume general arms orientations.}
	\label{fig:toymodel}
\end{figure}

In this work, we will consider a toy model for a Michelson-Morley interferometry experiment used to detect GWs. A pictorial description of such a model is represented on the first panel of figure \ref{fig:toymodel}\;. There, light is emitted from a laser $\mathcal{L}$ to a beam splitter $\mathcal{S}$ which separates the original signal into two equal-power halves. These, in turn, propagate to the end mirrors $\mathcal{M}$ and $\mathcal{M}'$ of their correspondent arms, where they are reflected back to the beam splitter and finally recombine at the photo-detector $\mathcal{P}$, providing a time-dependent interference pattern for light. 

This is a toy model due to several aspects. First, most of the technological elements of a real GW detector are neglected here, in particular, the usual Fabry-P\'{e}rot cavities, since interesting aspects are already apparent without them and the calculations are much simplified. Second, the distances between $\mathcal{L}$ and $\mathcal{S}$ and between $\mathcal{S}$ and $\mathcal{P}$ will be neglected, since these are small when compared with the arm's length. With this last observation in mind, the second panel of figure \ref{fig:toymodel} expresses the 4-dimensional description of the experiment, with observer $\mathcal{S}$ sending, in different instants, the beams that will interfere at the final event on this same observer. A third idealized characteristic of our model is the point-like nature of the mirrors. In this case, for an interferometry experiment to be successful, the light rays must be sent in a very restrictive way from $\mathcal{S}$ to each of the mirrors. To that end, a set of partial boundary conditions must be imposed on the light rays traveling each arm, which, together with the choice of a constant initial frequency value of the emitted rays, result in what we referred to as mixed conditions in L1. All the results of that first part related with the imposition of such conditions will be, then, valid here as well. A fourth and final aspect of the idealization worth mentioning is the assumption that light propagates as a test field, not generating any relevant curvature despite being affected by the presence of GWs.

Each constituent of the interferometer will be co-moving with the TT frame. We shall assume that observers $\mathcal{S}$, $\mathcal{M}$ and $\mathcal{M}'$ have arbitrary spatial coordinates $x^i_{\mathcal{S}}$, $x^i_{\mathcal{M}}$ and $x^i_{\mathcal{M}'}$, respectively, and consequently the arms are not necessarily orthogonal. Furthermore, as indicated on the second panel of figure \ref{fig:toymodel}, a particular ray outgoing from $\mathcal{S}$ into $\mathcal{M}$ will be called ray 1, while a ray incoming from $\mathcal{M}$ to $\mathcal{S}$ will be called ray 2. Analogously, ray 3 is emitted from $\mathcal{S}$ to $\mathcal{M}'$ and ray 4 from $\mathcal{M}'$ to $\mathcal{S}$. These rays are uniquely determined once the above mentioned mixed conditions are imposed (as in L1). Any quantity $Q$ defined in a given ray is denoted with a subscript $j \in \{1, 2, 3, 4\}$, that is, $Q_{|j}$. Events of emission and reflection on the arm ending on $\mathcal{M}$ are, respectively, $\mathcal{E}$ and $\mathcal{R}$, while for the arm ending on $\mathcal{M}'$ the corresponding events are $\mathcal{E}'$ and $\mathcal{R}'$. The common detection event is labeled as $\mathcal{D}$. Our ultimate goal is to obtain the final interference pattern, and, thus, we shall propagate the electric field using eq.~(\ref{eq:electric_evolution}) along rays 1, 2, 3 and 4 up to the event $\mathcal{D}$.

\section{The kinematics of the TT reference frame and the optical parameters} \label{sec:kinematics}

\subsection{TT frame kinematics and the optical parameters}

A reference frame, conceived as a congruence of observers \cite{Sachs1977}, has its local kinematics characterized, in analogy with a fluid in continuum mechanics, by the irreducible decomposition of its gradient:
\begin{equation}
	\nabla_{\beta} u_{\alpha} = - a_{\alpha} u_{\beta} + \frac{1}{3}\Theta p_{\alpha \beta} + \sigma_{\alpha \beta} + \Omega_{\alpha \beta}\,, \label{irrdecomp}
\end{equation}
where $p_{\alpha \beta} \leqdef g_{\alpha \beta} + u_{\alpha} u_{\beta}$ is the projector onto the local rest space: the 3-dimensional space orthogonal to $u^{\alpha}$ at each event. The expansion scalar $\Theta$, the shear tensor $\sigma_{\alpha \beta}$ and the vorticity tensor $\Omega_{\alpha \beta}$ describe the local relative motion among observers of the frame. Together with the 4-acceleration $a^{\alpha}$, we refer to them as the kinematic quantities (or parameters) of the reference frame $u^\alpha$ \cite{Ellis2012, Ehlers1961, Ehlers1993}.

For the case in which the metric components are (\ref{eq:metric}) and the TT reference frame $u^{\alpha} = \delta^{\alpha}_t$ is assumed, we find, from eq.~($\ref{gamma^t}$):  
\begin{equation}
	\nabla_{\beta} u_{\alpha} = \Gamma^t_{\beta \alpha} = \frac{1}{2}\epsilon_P h^P_{\beta \alpha,t}\,. \label{eq:nablau}
\end{equation}
This implies the following kinematic quantities:
\begin{equation}
	\Omega_{\alpha \beta} \leqdef p^{\gamma}_{\;[\beta} p^{\delta}_{\; \alpha]} \nabla_{\gamma} u_{\delta} = 0\,,
	\label{Omega}
\end{equation}
\begin{equation}
	\Theta \leqdef g^{\beta \alpha} p^{\gamma}_{\; \beta} p^{\delta}_{\; \alpha} \nabla_{\gamma} u_{\delta} = 0\,,
	\label{Theta}
\end{equation}
\begin{align}
	a_{\alpha} & \leqdef  u^{\beta}\nabla_{\beta}u_{\alpha} = 0\,,
	\label{a} \\
	\sigma_{\alpha \beta} & \leqdef p^{\gamma}_{(\beta} p^{\delta}_{\alpha)} \nabla_{\gamma} u_{\delta} - \frac{\Theta p_{\alpha \beta}}{3} = \frac{1}{2} \epsilon_P h^P_{\alpha \beta,t}\,,
	\label{Sigma}
\end{align}
from which we conclude that the TT frame has a purely shearing kinematics ($\nabla_{\alpha} u_{\beta} = \sigma_{\alpha \beta}$) induced by the presence of GWs. The usual heuristic infinitesimal (arm length much smaller than GW wavelength) description regarding interferometer response to a GW is closely attached to this kinematic property of the TT frame, to which the interferometer is assumed to be adapted. The detector deforms in an anisotropic way, such that the area of the rectangle defined by $\mathcal{S}$, $\mathcal{M}$ and $\mathcal{M}'$ being three of its vertices is preserved. This is the expected non-trivial contribution to the RHS of eq.~(\ref{eq:electric_evolution}), even when the interferometer is not in the long wavelength limit.

One may also define the so-called optical quantities $\widehat{a}^{\alpha}$, $\widehat{\Theta}$, $\widehat{\sigma}_{\alpha \beta}$  and $\widehat{\Omega}_{\alpha \beta}$, which are the kinematic parameters analogues for the description of a null congruence of curves whose tangent vectors are $k^{\alpha}$ \cite{Ellis2012}. Their definition is similar to the ones appearing in eqs.~(\ref{Omega})--(\ref{Sigma}), changing $u^{\alpha}$ for $k^{\alpha}$ and $p^{\alpha}_{\;\beta}$ for $s^{\alpha}_{\;\beta}$, where
\begin{equation}
s^{\alpha}_{\;\beta} \leqdef p^{\alpha}_{\;\beta} - n^{\alpha} n_{\beta}\,,	
\end{equation} 
is the projector onto the local screen space: the 2-dimensional space orthogonal to the instantaneous observer $u^{\alpha}$ and to the unit vector in the spatial direction along which light propagates, $n^{\alpha} \leqdef p^{\alpha}_{\;\beta} k^{\beta}/(-u^{\mu} k_{\mu})$ (equivalently, a vector in the screen space is orthogonal to both $u^\alpha$ and $k^\alpha$). 

Here we are concerned with the description of light rays,  working under the geometrical optics approximation where, for a given scalar function $\psi$,
\begin{equation}
	k_{\mu} = \nabla_{\mu} \psi\,,
\end{equation}
so that the null curves defining the congruence are geodesics and, by the analogous of eq.~(\ref{Omega}), $\widehat{\Omega}_{\alpha \beta} = 0$. Furthermore, the Faraday tensor in this regime reads \cite{Santana2020}:
\begin{equation}
	F_{\mu \nu} = \frac{(k_{\mu} E_{\nu} - k_{\nu} E_{\mu})}{\omega_{\textrm{e}}}\,,
\end{equation}
where $E^{\mu} \leqdef F^{\mu \nu} u_{\nu}$ is the electric field measured by $u^{\mu}$. An equivalent assertion is that $F_{\mu \nu}$ is a null bivector \cite{Hall2004}, namely, it obeys:
\begin{equation}
	F_{\mu \nu}F^{\mu \nu} = 0 = \frac{1}{2}\eta^{\alpha \beta \mu \nu} F_{\alpha \beta} F_{\mu \nu}\,, 
	\label{nullfield}
\end{equation} 
implying, finally, that $\widehat{\sigma}_{\alpha \beta} = 0$, if $F^{\mu \nu}$ satisfies Maxwell equations in vacuum \cite{Robinson1961}. 

The optical expansion $\widehat{\Theta} \leqdef {k^{\mu}}_{;\mu}$ is then the only non-vanishing optical parameter when light rays travel in vacuum following the usual laws of electrodynamics. It gives the divergence ($\widehat{\Theta}>0$) or convergence ($\widehat{\Theta}<0$) property of a beam of light (cf. eq.~(\ref{Thetaandarea})) and its evolution \cite{Ellis2012} along any of the rays simplifies to
\begin{equation}
	\frac{d\widehat{\Theta}}{d \vartheta} + \frac{1}{2} \widehat{\Theta}^2 = 0 \label{Thetaevol}
\end{equation}  
when the spacetime is Ricci-flat and no electromagnetic 4-current is present.

\subsection{Solving for the optical expansion along a ray}

The remaining ingredient for us to solve eq.~(\ref{eq:electric_evolution}) is the optical expansion $\widehat{\Theta}$ along a chosen light ray. From eq.~(\ref{Thetaevol}):
\begin{equation}
	\widehat{\Theta}_{|j}(\vartheta) = \frac{2\widehat{\Theta}_{|j}(0)}{2 + \vartheta \widehat{\Theta}_{|j}(0)}\,, \label{Thetasol}
\end{equation}
for $j = 1, 2, 3, 4$. We notice that, choosing $\vartheta > 0$ along the curve, if initially the beam is divergent, it will remain in this way, as one would expect, since eq.~(\ref{Thetaevol}) is also valid in Minkowski spacetime. Assuming a divergent beam from now on, and noting that $\widehat{\Theta}/2$ is the integrating factor of eq.~(\ref{eq:electric_evolution}), we find, for ray 1:
\begin{equation}
\textrm{e}^{-\frac{1}{2}\int_{0}^{\vartheta_{\mathcal{R}}} \widehat{\Theta}_{|1} d\vartheta} = \frac{1}{1+\frac{\vartheta_{\mathcal{R}}\widehat{\Theta}_{\mathcal{E}} }{2}}, \label{eq:exptheta1}
\end{equation}
where $\widehat{\Theta}_{\mathcal{E}} \leqdef \widehat{\Theta}_{|1}(0)$. A similar expression is valid for ray 2, being only necessary to make $\vartheta_{\mathcal{R}} \rightarrow \vartheta_{\mathcal{D}}$, $\widehat{\Theta}_{|1} \rightarrow \widehat{\Theta}_{|2}$ and $\widehat{\Theta}_{\mathcal{E}} \rightarrow  \widehat{\Theta}_{|2}(0)$.

Here we shall restrict to the case in which the optical expansion is continuously connected from ray 1 to ray 2, namely,
\begin{equation}
\widehat{\Theta}_{|2}(0) = \widehat{\Theta}_{|1}(\vartheta_{\mathcal{R}}). \label{contTheta}
\end{equation}
Imposing this initial condition for ray 2 in eq.~(\ref{Thetasol}), 
\begin{equation}
\textrm{e}^{-\frac{1}{2}\int_{0}^{\vartheta_{\mathcal{D}}} \widehat{\Theta}_{|2} d\vartheta} = \frac{1}{1+\frac{\vartheta_{\mathcal{D}}\widehat{\Theta}_{\mathcal{E}} }{2+\vartheta_{\mathcal{R}} \widehat{\Theta}_{\mathcal{E}}}}\,. \label{eq:exptheta2final}
\end{equation}

Note that this continuity assumption is only valid once a passive reflection in event $\mathcal{R}$ is guaranteed by means of a plane mirror. If the mirror was concave, $\widehat{\Theta}$ would have an abrupt change after reflection from a positive to a negative value; if it was convex, the discontinuity would still occur, although $\widehat{\Theta}$ would remain positive. The passive reflection can be contrasted with what occurs in a LISA-like interferometer \cite{Danzmann2017}. Since the arms in LISA are huge, the intensity of the emitted light dissipates and only a small amount reaches the other extremity. It is then necessary to generate a new, but phase-locked light beam at event $\mathcal{R}$ so that the final intensity pattern can be substantial in magnitude. Condition (\ref{contTheta}) cannot be achieved in such a framework, where one would more naturally have to impose
$\widehat{\Theta}_{|2}(0) = \widehat{\Theta}_{|1}(0)\,$.

To find $\widehat{\Theta}_{\mathcal{E}}$, we first relate the optical expansion with the beam's cross section area \cite{Ellis2012} (notice the typo in eq.~(7.25) therein):
\begin{equation}
\widehat{\Theta} = \frac{1}{\delta S}\frac{d}{d\vartheta}(\delta S)\,. \label{Thetaandarea}
\end{equation}
On the aLIGO experiment, the beam containing ray 1 has, to zeroth order in $\bm{\epsilon}$, a circular cross section of radius $5.3$ cm at the beginning of the arm and $6.2$ cm at its end \cite{Aasi2015, Martynov2016}. The corresponding cross sectional areas on the extremities are, then:
\begin{align}
	\delta S(0) & = 88 \,\text{cm}^2\,, \label{initarea} \\ 
	\delta S(\vartheta_{\mathcal{R}}) & = 121 \,\text{cm}^2\,. \label{finalarea}
\end{align} 
Because of the small dimension of the laser, we will assume that the initial condition for the optical expansion is not affected by GWs, \emph{i.e} it does not depend on $\bm{\epsilon}$. Integrating (\ref{Thetaandarea}) from $0$ to $\vartheta_{\mathcal{R}}$, inserting (\ref{initarea}), (\ref{finalarea}) and evaluating the LHS integral by (\ref{Thetasol}) one concludes that:
\begin{equation}
	\widehat{\Theta}_{\mathcal{E}} = \widehat{\Theta}_{(\bm{0})|1}(0) =  \frac{0.345}{{\vartheta_{\mathcal{R}}}_{(\bm{0})}} = 0.52 \times 10^3 \; \text{m}^{-2}, \label{eq:init_Theta_est}
\end{equation}
since (cf. L1) 
\begin{equation}
	{\vartheta_{\mathcal{R}}}_{(\bm{0})} = \frac{c \Delta \ell}{\omega_{e \mathcal{E}}}\,,
\end{equation}
where $\Delta \ell = 4$ km is the zeroth order aLIGO arm length and $\omega_{e \mathcal{E}} = 1.8 \times 10^{15}$ rad/s is the typical initial angular frequency at event $\cal{E}$ of an infrared laser in a LIGO-like experiment \cite{Martynov2016}. Eq.~(\ref{eq:init_Theta_est}) will be useful in estimating final interference pattern contributions related to intensity dissipation due to the beam divergence on the aLIGO case.

Of course the procedure developed in this subsection can be trivially extended to the other arm by changing ray 1 to 3 and 2 to 4 and the corresponding emission and reflection events (see figure \ref{fig:toymodel}). Also, our modeling of the experiment assumes hereafter that the initial frequency and optical expansion remain the same throughout the emission events along $\mathcal{S}$, in particular, although $\mathcal{E'} \neq \mathcal{E}$
\begin{align}
	\omega_{e \mathcal{E'}} = \omega_{e \mathcal{E}}\,, \quad
	\widehat{\Theta}_{\mathcal{E'}} = 	\widehat{\Theta}_{\mathcal{E}}\,.
\end{align}
\section{The electric field evolution in the geometric optics limit} \label{sec:electriceq}
Here we reserve a space to discuss the physical significance of eq.~(\ref{eq:electric_evolution}) for the propagation of electric fields along light rays  under the geometrical optics limit on arbitrary spacetimes when these are measured by a general reference frame whose field of instantaneous observers is given by $u^{\mu}$. 

Expressing $E^{\mu} \reqdef E e^{\mu}$, where $E \leqdef \sqrt{E^{\mu} E_{\mu}}$, we may dismember such equation into two parts \cite{Santana2020}:
\begin{equation}
\frac{De^{\mu}}{d\vartheta} = k^{\mu} \frac{e^{\nu}}{\omega_{\textrm{e}}} \frac{Du_{\nu}}{d\vartheta}\,,
\label{polevol}
\end{equation}
and
\begin{equation}
\frac{dE}{d\vartheta} + \frac{\widehat{\Theta}}{2}E = -\frac{k^{\nu}E}{\omega_{\textrm{e}}}\frac{Du_{\nu}}{d\vartheta} = \frac{E}{\omega_{\textrm{e}}}\frac{d\omega_{\textrm{e}}}{d\vartheta}\,. \label{Magevol}
\end{equation}

From eq.~(\ref{polevol}), one concludes that the first term of the RHS of eq.~(\ref{eq:electric_evolution}) is present if and only if the electric field polarization $e^{\mu}$ is not parallel transported along the light ray. For a Faraday tensor satisfying eq.~(\ref{nullfield}) the (instantaneous) intensity of light is given by
\begin{equation}
	I = g_{\mu\nu} E^\mu E^\nu\,,
\end{equation}
and from eq.~(\ref{Magevol}) it is easy to deduce how it propagates, namely
\begin{equation}
	\frac{dI}{d\vartheta} + \widehat{\Theta} I = \frac{2I}{\omega_{\textrm{e}}}\frac{d\omega_{\textrm{e}}}{d\vartheta}\,,
	\label{eq:intensity_evolution}
\end{equation}
from which, together with eq.~(\ref{Thetaandarea}), we obtain
\begin{equation}
	\frac{I\delta S}{\omega_{\textrm{e}}^2} = \text{const}
	\label{eq:brightness_conservation}
\end{equation}
along the chosen geodesic, which stands for the conservation of photon number of the light beam \cite{Schneider1992}.

The intimate relation between eqs.~(\ref{Magevol}) and (\ref{eq:intensity_evolution}) allows us to interpret the physical origins of the other terms present in eq.~(\ref{eq:electric_evolution}). In eq.~(\ref{Magevol}), the RHS of each equality demonstrates that the second term on the RHS of eq.~(\ref{eq:electric_evolution}) is a consequence of a frequency shift effect arising from the kinematics of the reference frame. This contribution is expected not to vanish as demonstrated in L1. The physical interpretation surrounding it is clear once one notices that the energy of each photon is proportional to its frequency and that, as a result, a shift in $\omega_{\textrm{e}}$ should alter the intensity of light, which is confirmed by eq.~(\ref{eq:brightness_conservation}). Finally, the term proportional to $\widehat{\Theta}$ accounts for the increase or decrease in intensity following a possible convergence or divergence of the beam, respectively.

In order to compute $Du^{\mu}/d\vartheta$ in eq.(\ref{eq:electric_evolution}), although it is only necessary to define a differentiable vector field of instantaneous observers on the light ray of interest, one needs an entire reference frame if the electric field is to be propagated along all possible rays continuously emitted for a general interferometer configuration. In this case, such an absolute derivative can be expressed in terms of the kinematics of the given frame by:
\begin{equation}
	\frac{Du^{\mu}}{d\vartheta} = k^{\nu} \nabla_{\nu} u^{\mu}\,.
\end{equation}   
We assume, then, the presence of the TT frame on an open set containing all possible interferometer orientations. As a consequence, we highlight that it is the shearing character of this frame that induces the non-parallel transport of the polarization vector and the frequency shift of light, which contribute to a non-trivial propagation of $E^{\mu}$.

One could try to use directly eq.~(\ref{eq:intensity_evolution}) to propagate the intensity in each ray and calculate the final interference pattern of the experiment. But, since we allow the polarization vector of rays 2 and 4 on event $\mathcal{D}$ to be differently perturbed by the GW, the relation between the final intensity on each of these two rays before interference with the total intensity after it is not obvious. Because of this, we choose to solve eq.~(\ref{eq:electric_evolution}) instead.  
\section{Electric field solution} \label{sec:electricsolutions}

On this section we aim to solve eq.~(\ref{eq:electric_evolution}), using the metric (\ref{eq:metric}) and the TT comoving observers $u^{\alpha} = \delta^{\alpha}_t$, along the null geodesics connecting them. First, since $g_{ti} = 0$,
\begin{equation}
	E^t \leqdef F^{t \nu} u_{\nu} = 0,
\end{equation} 
and thus the non-trivial part of eq.~(\ref{eq:electric_evolution}) becomes the spatial one. Taking into account eq.~(\ref{eq:nablau}) and that, as presented in appendix \ref{app:christoffel_symbols}, all Christoffel symbols are of order $\bm{\epsilon}$, it can be conveniently rewritten as:
\begin{equation}
	\frac{dE^{i}}{d\vartheta} + \frac{1}{2} \widehat{\Theta} E^{i} = f^i, \label{eq:electricpropGW}
\end{equation}
with
\begin{align}
	f^i(\vartheta) \leqdef k^{\beta}_{(\bm{0})} \bigg[\bigg(  \frac{\zeta k^{i}_{(\bm{0})} E^{\nu}_{(\bm{0})}(\vartheta) - k^{\nu}_{(\bm{0})} E^{i}_{(\bm{0})}(\vartheta)}{\omega_{e(\bm{0})}}\bigg) \Gamma^t_{\beta \nu} (\vartheta)   \nonumber - \Gamma^{i}_{\beta j} (\vartheta) E^j_{(\bm{0})}(\vartheta) \bigg], \label{eq:f}
\end{align}
where the last term arises from the absolute derivative of the electric field. The quantity $\zeta = 0,1$ was introduced by hand to monitor the presence of the contribution related to the RHS of eq.~(\ref{polevol}) and allows us to conclude whether the non-parallel transport of $e^{\mu}$ induced by the TT frame kinematics will result in non-trivial additional contributions to the electric field along each ray and, if so, to the intensity pattern.  

Such a propagation equation is then solved perturbatively as follows. To zeroth order, eq.~(\ref{eq:electricpropGW}) becomes
\begin{equation}
	\frac{dE^{i}_{(\bm{0})}}{d\vartheta} + \frac{1}{2} \widehat{\Theta}_{(\bm{0})}E^{i}_{(\bm{0})} = 0,
\end{equation}
whose solution is simply
\begin{equation}
	E^i_{(\bm{0})}(\vartheta) = E^i_{(\bm{0})}(0) \textrm{e}^{-\frac{1}{2}\int_{0}^{\vartheta}\hat{\Theta}_{(\bm{0})}(\vartheta')d\vartheta'}. \label{eq:zerothorder}
\end{equation}
As for the linear order solution, one can easily integrate eq.~(\ref{eq:electricpropGW}) with the help of an integrating factor. Once eq.~(\ref{eq:zerothorder}) is replaced on such a solution, one finds: 
\begin{align}
		E^i(\vartheta) = \textrm{e}^{-\frac{1}{2}\int_{0}^{\vartheta}\hat{\Theta}(\vartheta')d\vartheta'} \bigg\{E^i(0) \nonumber + k^{\beta}_{(\bm{0})}  \bigg[\frac{[\zeta k^{i}_{(\bm{0})} E^{\nu}_{(\bm{0})}(0) - k^{\nu}_{(\bm{0})} E^{i}_{(\bm{0})}(0)]}{\omega_{\textrm{e}\mathcal{E}}}  &\int_{0}^{\vartheta} \Gamma^t_{\beta \nu} (\vartheta') d\vartheta'  \nonumber \\ - E^{j}_{(\bm{0})}(0) &\int_{0}^{\vartheta} \Gamma^{i}_{\beta j}(\vartheta') d\vartheta' \bigg]   \bigg\}\,, \label{eq:electricmidway}
\end{align}
where we have assumed that the laser frequency at emission is constant along the source $\mathcal{S}$ and not perturbed by GWs, so that $\omega_{\textrm{e}(\bm{0})} = \omega_{\textrm{e}\mathcal{E}} = \omega_{\textrm{e}\mathcal{E'}}$.
Until now, the GW amplitude was thought of as a function of $t-x$. Its restriction to each ray is
\begin{equation}
	h_{P|j}(\vartheta) \leqdef h_P (\xi^t_{|j}(\vartheta) - \xi^x_{|j}(\vartheta))\,,
\end{equation}
where $\xi^{\alpha}_{|j}(\vartheta)=x^{\alpha}$ is the $\alpha$ component of the parametrized ray $j$. The family of possible null geodesics obtained in L1 satisfies
\begin{equation}
\xi^u(\vartheta) \leqdef \frac{\xi^t(\vartheta) - \xi^x(\vartheta)}{2} = -\vartheta \delta + \xi^u(0)\,,
\end{equation} 
where $\delta$ is a constant of motion.
The first integral in eq.~(\ref{eq:electricmidway}), together with eq.~(\ref{gamma^t}), can then be solved:
\begin{align}
	\int_{0}^{\vartheta} \Gamma^{t}_{\beta \nu}(\vartheta') d\vartheta' = 	\frac{\epsilon_P}{4}\int_{0}^{\vartheta} h^{P}_{\beta \nu,u}(\vartheta') d\vartheta'  & = \frac{\epsilon_P}{4}\int_{0}^{\vartheta} \frac{dh^{P}_{\beta \nu}}{d\vartheta'} \bigg(\frac{d\xi^u}{d\vartheta'}\bigg)^{-1} d\vartheta' \nonumber \\ & =-\frac{\epsilon_P}{4 \delta} [h^P_{\beta \nu}(\vartheta) - h^P_{\beta \nu}(0)]\,. \label{eq:inth}
\end{align}
The remaining integrals are analogously handled with the help of eqs.~($\ref{gamma_t}$)--($\ref{gamma_z}$).

We now particularize to ray 1. From the results of L1, after imposing that the ray leaves $\mathcal{S}$ and reaches $\mathcal{M}$:
\begin{equation}
	\frac{1}{\delta_{|1}} = \frac{\Delta \ell}{\omega_{\textrm{e}\mathcal{E}}(\Delta x - \Delta \ell)}  + \mathcal{O}(\bm{\epsilon})\,, \label{eq:delta}
\end{equation}
where $\Delta x^i \leqdef x^i_{\mathcal{M}}- x^i_{\mathcal{S}}$ and $\Delta \ell \leqdef \sqrt{\delta_{ij} \Delta x^i \Delta x^j}$. Under the same conditions, the null geodesics in Minkowski spacetime obey:
\begin{equation}
	k^{\alpha}_{\bm{(0)}} = \omega_{\textrm{e}\mathcal{E}} \bigg(\delta^{\alpha}_t + \frac{\Delta x^i}{\Delta \ell} \delta^{\alpha}_i \bigg). \label{eq:kMink}
\end{equation}
Evaluating eq.~($\ref{eq:electricmidway}$) on ray 1, replacing the solved Christoffel symbol integrals together with eqs.~($\ref{eq:delta}$) and ($\ref{eq:kMink}$), and recalling eqs.~($\ref{eq:hplus}$) and ($\ref{eq:hcross}$), one concludes that: 
\begin{equation}
E^i_{|1}(\vartheta) = \textrm{e}^{-\frac{1}{2}\int_{0}^{\vartheta}\widehat{\Theta}_{|1}(\vartheta')d\vartheta'} [E^i_{|1}(0) + J^i_{|1}(\vartheta)]\,, \label{eq:Eray1}
\end{equation}
where:
\begin{align}
	&\hspace{-2pt} J^x_{|1}(\vartheta) = \frac{1}{2(\Delta \ell-\Delta x)} \times \nonumber \\ & \hspace{40pt} \times \bigg\{ \epsilon_+ \Delta h_{+|1}(\vartheta) \bigg[\bigg( 1 - \zeta \frac{\Delta x}{\Delta \ell}\bigg) \big(\Delta y E^y_{(\bm{0})|1}(0) -\Delta z E^z_{(\bm{0})|1}(0) \big) + \frac{E^x_{(\bm{0})|1}(0)}{\Delta \ell}\big(\Delta y^2 -\Delta z^2\big) \bigg]  \nonumber \\ & \hspace{48pt}+ \epsilon_{\times} \Delta h_{\times|1}(\vartheta) \bigg[ \bigg( 1 - \zeta \frac{\Delta x}{\Delta \ell}  \bigg) (\Delta y E^z_{(\bm{0})|1}(0) + \Delta z E^y_{(\bm{0})|1}(0)) + 2\frac{\Delta z \Delta y}{\Delta \ell}E^x_{(\bm{0})|1}(0) \bigg]\bigg\}, \label{M1}
\end{align}
\begin{align}
	&J^y_{|1}(\vartheta) = \frac{1}{2(\Delta \ell-\Delta x)} \bigg\{ \epsilon_+ \Delta h_{+|1}(\vartheta)\bigg[ E^z_{(\bm{0})|1}(0)\frac{\zeta \Delta z \Delta y}{\Delta \ell} - E^x_{(\bm{0})|1}(0) \Delta y  \nonumber \\ & \hspace{209pt} + E^y_{(\bm{0})|1}(0)\bigg(\Delta \ell - \Delta x + \frac{\Delta y^2(1 - \zeta) - \Delta z^2}{\Delta \ell}\bigg) \bigg] \nonumber \\ &\hspace{20pt} + \epsilon_{\times} \Delta  h_{\times|1}(\vartheta)\bigg[ E^y_{(\bm{0})|1}(0)\frac{\Delta y \Delta z}{\Delta \ell}(2 - \zeta) - E^x_{(\bm{0})|1}(0)\Delta z  + E^z_{(\bm{0})|1}(0)\bigg( \Delta \ell - \Delta x - \frac{\zeta \Delta y^2}{\Delta \ell} \bigg) \bigg]\bigg\}\label{M2},
\end{align}
\begin{align}
	&\hspace{-4pt}J^z_{|1}(\vartheta) = \frac{1}{2(\Delta \ell-\Delta x)}\bigg\{ \epsilon_+\Delta h_{+|1}(\vartheta)\bigg[E^x_{(\bm{0})|1}(0)\Delta z  - E^y_{(\bm{0})|1}(0)\frac{\zeta \Delta z \Delta y}{\Delta \ell} \nonumber \\
	& \hspace{204pt} + E^z_{(\bm{0})|1}(0)\bigg( \Delta x - \Delta \ell + \frac{\Delta z^2(\zeta - 1) + \Delta y^2}{\Delta \ell}\bigg)  \bigg] \nonumber \\
	& \hspace{15pt}+ \epsilon_{\times}\Delta h_{\times|1}(\vartheta)\bigg[ E^z_{(\bm{0})|1}(0)\frac{\Delta y \Delta z}{\Delta \ell}(2 - \zeta) - E^x_{(\bm{0})|1}(0) \Delta y + E^y_{(\bm{0})|1}(0)\bigg( \Delta \ell - \Delta x - \frac{\zeta \Delta z^2}{\Delta \ell} \bigg) \bigg]\bigg\}. \label{eq:M3}
\end{align}
and $\Delta h_{P|j}(\vartheta) \leqdef h_{P|j}(\vartheta) - h_{P|j}(0)$. 

With the aid of figure \ref{fig:toymodel}, it is easy to see that the field on ray 3 is obtained from eqs.~(\ref{eq:Eray1})--(\ref{eq:M3}) by changing $\Delta x^i \rightarrow \Delta x'^{i} \leqdef x_{\mathcal{M'}} - x_{\mathcal{S}}$, $\Delta \ell \rightarrow \Delta \ell' \leqdef \sqrt{\delta_{ij}\Delta x'^{i}\Delta x'^{j}}$, $\widehat{\Theta}_{|1} \rightarrow \widehat{\Theta}_{|3}$, $E^i_{|1}(0) \rightarrow  E^i_{|3}(0)$ and $\Delta h_{P|1} \rightarrow \Delta h_{P|3}$. On the other hand, the field on ray 2 is a consequence of making the changes $\Delta x^i \rightarrow -\Delta x^{i}$, $\widehat{\Theta}_{|1} \rightarrow \widehat{\Theta}_{|2}$, $\Delta h_{P|1} \rightarrow \Delta h_{P|2}$ and $E^i_{|1}(0) \rightarrow E^i_{|2}(0) = -E^i_{|1}(\vartheta_{\mathcal{R}})$; the latter change expressing the usual phase shift of $\pi$ when light is assumed to be reflected by a perfect mirror. Finally, for ray 4, one should do the same changes when passing from ray 1 to 2, but on the expression of $E^i_{|3}(\vartheta)$.
\section{Intensity pattern for GW normal incidence} \label{sec:finalintensity}
To calculate the final intensity pattern on our toy model interferometer, the fields coming from rays 2 and 4 must be evaluated at event $\mathcal{D}$ (cf. figure \ref{fig:toymodel}). From the results of the preceding section, $E^i_{|2}(\vartheta_{\mathcal{D}})$ can be recast in terms of the initial value of the electric field $E^i_{|1}(0)$, the GW amplitude $h_P$ and the known quantities $\Delta x^i$, $\omega_{\textrm{e}\mathcal{E}}$ and $\widehat{\Theta}_{\mathcal{E}}$. To that end, it suffices to notice that, from the analogue of eq.~(\ref{eq:Eray1}) for ray 2
\begin{align}
	E^i_{|2}(\vartheta_{\mathcal{D}}) &= \textrm{e}^{-\frac{1}{2}\int_{0}^{\vartheta_{\mathcal{D}}}\widehat{\Theta}_{|2}(\vartheta')d\vartheta'} [-E^i_{|1}(\vartheta_{\mathcal{R}}) + J^i_{|2}(\vartheta_{\mathcal{D}})] \nonumber \\&= \textrm{e}^{-\frac{1}{2}\int_{0}^{\vartheta_{\mathcal{D}}}\widehat{\Theta}_{|2}(\vartheta')d\vartheta'} \Big\{-\textrm{e}^{-\frac{1}{2}\int_{0}^{\vartheta_{\mathcal{R}}}\widehat{\Theta}_{|1}(\vartheta')d\vartheta'} [E^i_{|1}(0)+J_{|1}^i(\vartheta_{\mathcal{R}})] + J^i_{|2}(\vartheta_{\mathcal{D}})\Big\}, \label{eq:Eray2}
\end{align}
where, in $J^i_{|2}(\vartheta_{\mathcal{D}})$, the initial field $E^i_{(\bm{0})|2}(0)$ can be expressed in terms of $E^i_{(\bm{0})|1}(0)$ as
\begin{equation}
	E^i_{(\bm{0})|2}(0) = - E^i_{(\bm{0})|1}(0) \textrm{e}^{-\frac{1}{2}\int_{0}^{\vartheta_{\mathcal{R}}}\widehat{\Theta}_{|1}(\vartheta')d\vartheta'}, \label{eq:initE2}
\end{equation}
so that the exponential terms can be factored out. Then, it is only necessary to replace eqs.~(\ref{eq:exptheta1}) and (\ref{eq:exptheta2final}) on eq.~(\ref{eq:Eray2}) together with the expressions of $\vartheta_{\mathcal{R}}$ and $\vartheta_{\mathcal{D}}$ calculated on subsection V A of L1. Here, to simplify our analysis, we calculate it in the particular case in which the incidence of the GW (which is traveling in the $x$ direction) is normal to the detector apparatus, namely, we consider $\Delta x = \Delta x' = 0$. 

Let an auxiliary ray, call it ray 5, be the one leaving event $\mathcal{E}$ together with ray 1, but on the other arm of the interferometer. These rays are the transmitted and reflected halves of the original laser beam, divided by the splitter. Immediately after this splitting, our starting event $\mathcal{E}$, their electric fields differ only by a minus sign due to the reflection
\begin{equation}
E^i_{|5}(0) = - E^i_{|1}(0)\,, \label{eq:refl_phase_shift}
\end{equation}
and, thus, imposing the geometrical optics result of the transversality of the electric field, in zeroth order, on both rays at their common emission, we have
\begin{equation}
k_{i (\bm{0})|1}(0)E^{i}_{(\bm{0})|1}(0) = - k_{i (\bm{0})|5}(0)E^{i}_{(\bm{0})|1}(0) = 0\,. \label{eq:transv_k_E}
\end{equation}
One concludes, therefore, in the case of normal GW incidence, that
\begin{equation}
E^y_{(\bm{0})|1}(0) = E^z_{(\bm{0})|1}(0) = 0\,. \label{eq:vanishing_unpert_comp}
\end{equation}
Here it is important to stress that such a result does not depend on the value of the angle between the arms of the interferometer, which is treated as arbitrary throughout both parts of this series of papers. It is only a consequence of the fact that, at zeroth order, the initial electric field must be orthogonal to the rays simultaneously leaving the beam splitter in both arms which define a plane that, in the normal incidence case, is orthogonal to the GW propagation direction, namely the $yz$ plane. An analogous argument is made to conclude that $E^y_{(\bm{0})|3}(0) = E^z_{(\bm{0})|3}(0) = 0\,.$

We emphasize that a general light ray orthogonal to the GW propagation direction in a more general physical situation does not necessarily have its zeroth order electric field parallel to the GW. That is the case for interferometers because of the beam splitter device, that connects the initial electric fields at each arm (eq.~($\ref{eq:refl_phase_shift}$)). Finally, because of eq.~($\ref{eq:zerothorder}$),
\begin{equation}
	 E^y_{(\bm{0})|j}(\vartheta) = E^z_{(\bm{0})|j}(\vartheta)=0. \label{eq:zeroth_electric_field_vanish}
\end{equation}

With all of this in mind, $J^i_{|1}$ and $J^i_{|2}$ simplify to
\begin{align}
	J^x_{|j}(\vartheta) & = \frac{1}{2\Delta \ell^2} [\epsilon_+ \Delta h_{+|j}(\vartheta)(\Delta y^2 - \Delta z^2) +  2\epsilon_{\times} \Delta h_{\times|j}(\vartheta) \Delta y \Delta z] E^x_{(\bm{0})|j}(0), \label{eq:Jxnorm} \\
	J^y_{|j}(\vartheta) &= \frac{(-1)^j}{2 \Delta \ell} [\epsilon_+ \Delta h_{+|j}(\vartheta)\Delta y + \epsilon_{\times} \Delta h_{\times|j}(\vartheta) \Delta z ] E^x_{(\bm{0})|j}(0), \label{eq:Jynorm} \\
	J^z_{|j}(\vartheta) &= \frac{(-1)^{j+1}}{2\Delta \ell} [\epsilon_+ \Delta h_{+|j}(\vartheta)\Delta z - \epsilon_{\times} \Delta h_{\times|j}(\vartheta) \Delta y ] E^x_{(\bm{0})|j}(0), \label{eq:Jznorm}
\end{align}
where $j=1,2$ in the above expressions. The exponential factors become 
\begin{align}
& \textrm{e}^{-\frac{1}{2}\left[\int_{0}^{\vartheta_{\mathcal{D}}}\widehat{\Theta}_{|2}(\vartheta')d\vartheta' +\int_{0}^{\vartheta_{\mathcal{R}}}\widehat{\Theta}_{|1}(\vartheta')d\vartheta'\right]} = \frac{1}{1+\frac{\widehat{\Theta}_{\mathcal{E}}(\vartheta_{\mathcal{R}}+ \vartheta_{\mathcal{D}})}{2}}  = \frac{\omega_{\textrm{e}\mathcal{E}}}{\widehat{\Theta}_{\mathcal{E}}\Delta \ell + \omega_{\textrm{e}\mathcal{E}}} \bigg \{1 - \frac{\widehat{\Theta}_{\mathcal{E}}}{(\widehat{\Theta}_{\mathcal{E}}\Delta \ell + \omega_{\textrm{e}\mathcal{E}})} \times \nonumber \\ & \hspace{10pt} \times \bigg[2\Delta L(t_{\mathcal{D}}- \Delta \ell) + \epsilon_{+} \frac{\Delta y^2 - \Delta z^2}{2\Delta \ell} h_{+}(t_{\mathcal{D}} - 2 \Delta \ell - x_{\mathcal{S}}) + \epsilon_{\times}\frac{\Delta y \Delta z}{\Delta \ell} h_{\times}(t_{\mathcal{D}} - 2 \Delta \ell - x_{\mathcal{S}})\bigg]\bigg\}, \label{eq:expnorm}
\end{align}
where $\Delta L(t) \leqdef D_R(t) - \Delta \ell$ is the radar distance perturbation presented in L1, for normal incidence: 
\begin{align}
	\Delta L (t) =  -\frac{1}{2} \left\{\epsilon_+ \frac{\Delta y^2 - \Delta z^2}{2\Delta \ell^2} \int^{t+\Delta \ell -x_{\mathcal{S}}}_{t-\Delta \ell -x_{\mathcal{S}}} h_+(w) dw + \epsilon_{\times} \frac{\Delta y \Delta z}{\Delta \ell^2} \int^{t+\Delta \ell -x_{\mathcal{S}}}_{t-\Delta \ell -x_{\mathcal{S}}} h_{\times}(w) dw \right\}. \label{eq:Delta_L}
\end{align}

As will be done with the final expressions of the electric field components, we chose to express the terms in  eq.~(\ref{eq:expnorm}) as functions of the detection time $t_{\mathcal{D}}$, using that:
\begin{align}
h_{P|2}(\vartheta_{\mathcal{D}}) &= h_P(t_{\mathcal{D}} - x_{\mathcal{S}}), \\
\epsilon_P h_{P|1}(\vartheta_{\mathcal{R}}) = \epsilon_P h_{P|2}(0) = \epsilon_P h_P(t_{\mathcal{R}} - x_{\mathcal{M}}) &= \epsilon_P h_P(t_{\mathcal{D}} - \Delta \ell - x_{\mathcal{M}}), \\
\epsilon_P h_{P|1}(0) = \epsilon_P h_P(t_{\mathcal{E}} - x_{\mathcal{S}}) &= \epsilon_P h_P(t_{\mathcal{D}} - 2\Delta \ell - x_{\mathcal{S}}).
\end{align}

Inserting eq.~($\ref{eq:initE2}$) in eqs.~($\ref{eq:Jxnorm}$)--($\ref{eq:Jznorm}$) and the result in eq.~($\ref{eq:Eray2}$) together with eq.~($\ref{eq:expnorm}$), the final electric field on ray 2, in the normal incidence case reads:
\begin{align}
E^x_{|2}(t_{\mathcal{D}}) & = - \frac{\omega_{\textrm{e}\mathcal{E}}E^x_{|1}(t_{\mathcal{E}})}{\widehat{\Theta}_{\mathcal{E}} \Delta \ell + \omega_{\textrm{e}\mathcal{E}}} \bigg\{1 + \frac{1}{\Delta \ell^2}\bigg[\frac{\Delta y^2 - \Delta z^2}{2} \epsilon_+ F_+(t_{\mathcal{D}}) + \Delta y \Delta z \epsilon_{\times }F_{\times}(t_{\mathcal{D}}) \bigg]  \bigg\}, \label{eq: x_electr_comp_norm_inc}\\ \nonumber \\
E^y_{|2}(t_{\mathcal{D}}) & = -\frac{\omega_{\textrm{e}\mathcal{E}}}{\widehat{\Theta}_{\mathcal{E}} \Delta \ell + \omega_{\textrm{e}\mathcal{E}}} \bigg\{E^y_{|1}(t_{\mathcal{E}}) +\frac{1}{2\Delta \ell}E^x_{(\bm{0})|1}(t_{\mathcal{E}})\bigg[\Delta y G_+(t_{\mathcal{D}}) + \Delta z G_{\times}(t_{\mathcal{D}})\bigg] \bigg\},\label{eq: y_electr_comp_norm_inc}\\ \nonumber \\
E^z_{|2}(t_{\mathcal{D}}) & = \frac{\omega_{\textrm{e}\mathcal{E}}}{\widehat{\Theta}_{\mathcal{E}} \Delta \ell + \omega_{\textrm{e}\mathcal{E}}} \bigg\{E^z_{|1}(t_{\mathcal{E}}) + \frac{1}{2\Delta \ell}E^x_{(\bm{0})|1}(t_{\mathcal{E}})\bigg[\Delta z G_+(t_{\mathcal{D}}) - \Delta y G_{\times}(t_{\mathcal{D}})\bigg] \bigg\}, \label{eq: z_electr_comp_norm_inc}
\end{align}
where
\begin{align}
F_{P}(t_{\mathcal{D}}) \leqdef & \, h_{P}(t_{\mathcal{D}} - x_{\mathcal{S}}) - h_{P}(t_{\mathcal{D}} - 2\Delta \ell - x_{\mathcal{S}}) \nonumber\\ & +\frac{\widehat{\Theta}_{\mathcal{E}}}{\widehat{\Theta}_{\mathcal{E}} \Delta \ell + \omega_{\textrm{e}\mathcal{E}}} \bigg[ \int^{t_{\mathcal{D}} -x_{\mathcal{S}}}_{t_{\mathcal{D}}-2\Delta \ell -x_{\mathcal{S}}} h_{P}(w) dw -\Delta \ell h_{P}(t_{\mathcal{D}} - 2\Delta \ell - x_{\mathcal{S}}) \bigg]
\end{align}
and
\begin{align}
G_P(t_{\mathcal{D}}) \leqdef\, h_{P}(t_{\mathcal{D}} - x_{\mathcal{S}}) - h_{P}(t_{\mathcal{D}} - \Delta \ell - x_{\mathcal{M}}) - [h_{P}(t_{\mathcal{D}} - \Delta \ell - x_{\mathcal{M}}) - h_{P}(t_{\mathcal{D}} - 2\Delta \ell - x_{\mathcal{S}})].
\end{align}

Here we notice the vanishing of the $\zeta$ contributions. This is expected in the case of normal incidence since, because of eq.~ ($\ref{eq:zeroth_electric_field_vanish}$), the unperturbed polarization vector is parallel to the GW propagation direction, i.e. $e^{\nu}_{\bm{(0)}|j} = e^x_{\bm{(0)}|j} \delta^{\nu}_{x} , \forall j$, and so it is orthogonal to the shear tensor of eq.~($\ref{Sigma}$), which implies eq.~($\ref{polevol}$) to take the form:
\begin{equation}
	 \frac{De^{\mu}_{|j}}{d\vartheta} =   k^{\mu}_{|j}k^{\alpha}_{|j} e^{\nu}_{|j} \sigma_{\alpha \nu}  = \frac{1}{2}\epsilon_P k^{\mu}_{|j}k^{\alpha}_{|j} e^{x}_{\bm{(0)}|j} h^P_{ \alpha x,t} = 0\,,
\end{equation}
and so the electric field polarization is indeed parallel transported. For arbitrary incidence, the shearing $yz$ plane will not coincide with the plane of the arms, allowing $e^{\nu}$ not to be perpendicular to $\sigma_{\alpha \beta}$, and so we expect $\zeta$ contributions not to vanish.

Under the same circumstances, $E^i_{|4}$ may be obtained by making $\Delta y \rightarrow \Delta y'$, $\Delta z \rightarrow \Delta z'$, $x_{\mathcal{M}} \rightarrow x_{\mathcal{M'}}$, $\Delta \ell \rightarrow \Delta \ell'$ and $E^i_{|1}(t_{\mathcal{E}}) \rightarrow E^i_{|3}(t_{\mathcal{E'}})$ in eqs.~($\ref{eq: x_electr_comp_norm_inc}$)--($\ref{eq: z_electr_comp_norm_inc}$). 

%The fact that $\zeta$ does not appear on (\ref{eq: x_electr_comp_norm_inc})--(\ref{eq: z_electr_comp_norm_inc}) must, then, be a consequence of the transversality impositions (\ref{eq:transv_k_E}). This is expected, since such tranversality leaves an unique degree of freedom for the zero order electric field, allowing one to think of $E^i_{(0)}$ as a scalar field. Since, on such field expressions, all terms of linear order in $\bm{\epsilon}$ depend solely on the zero order electric field components, the $\zeta$ contributions vanish as a consequence of this rationale. This argument suggests that even in a general GW incidence, the non-parallel transport law of the polarization does not affect the electric field propagation, at least under the perturbative scheme here adopted. 

Then, computing the total electric field at $\mathcal{D}$
\begin{equation}
E^{\mu}_T(t_{\mathcal{D}}) \leqdef E^{\mu}_{|2}(t_{\mathcal{D}}) +  E^{\mu}_{|4}(t_{\mathcal{D}}),
\end{equation} 
using eqs.~(\ref{eq:metric}) and (\ref{eq:zeroth_electric_field_vanish}), we find the final intensity:
\begin{align}
I(t_{\mathcal{D}}) &= [\eta_{\mu \nu} + \epsilon_P h^P_{\mu \nu}(t_{\mathcal{D}} - x_{\mathcal{S}})] E^{\mu}_T(t_{\mathcal{D}}) E^{\nu}_T(t_{\mathcal{D}}) = [E^x_T(t_{\mathcal{D}})]^2. \label{eq: final_intensity_simplification}
\end{align}

We then develop the usual procedure of relating the final electric fields to the distance traveled by each ray assuming a harmonic initial condition for the electric field:
\begin{align}
	E^x_{|1}(t_{\mathcal{E}}) = \mathcal{E}^x \cos(\omega_{\textrm{e}\mathcal{E}}t_{\mathcal{E}}) &= \mathcal{E}^x \cos[\omega_{\textrm{e}\mathcal{E}}(t_{\mathcal{D}} - 2D_{R}(t_{\mathcal{D}}- \Delta \ell))] \nonumber \\&= \mathcal{E}^x\{ \cos[\omega_{\textrm{e}\mathcal{E}}(t_{\mathcal{D}} - 2\Delta \ell)] + 2 \omega_{e \mathcal{E}} \Delta L(t_{\mathcal{D}} - \Delta \ell) \sin[\omega_{e \mathcal{E}}(t_{\mathcal{D}} - 2\Delta \ell)]\}\,, \label{eq: x_comp_init_cond_ray_1}
\end{align}
where $\mathcal{E}^x$ is a constant amplitude along $\mathcal{S}$ and an expansion on $\bm{\epsilon}$ was made in the last equality. For $E^x_{|3}(t_{\mathcal{E'}})$, we write:   
\begin{align}
E^x_{|3}(t_{\mathcal{E'}}) &= -\mathcal{E}^x \cos(\omega_{\textrm{e}\mathcal{E}}t_{\mathcal{E'}})  \nonumber \\ &= -\mathcal{E}^x\{ \cos[\omega_{\textrm{e}\mathcal{E}}(t_{\mathcal{D}} - 2\Delta \ell'] + 2 \omega_{e \mathcal{E}} \Delta L'(t_{\mathcal{D}} - \Delta \ell') \sin[\omega_{e \mathcal{E}}(t_{\mathcal{D}} - 2\Delta \ell')]\}\,, \label{eq: x_comp_init_cond_ray_3}
\end{align}
where the relative negative sign is, again, a consequence of the initial reflection on the beam-splitter and $D_R' \leqdef \Delta \ell' + \Delta L'$ is the radar distance of the other arm.

Replacing in eq.~(\ref{eq: final_intensity_simplification}), eq.~(\ref{eq: x_electr_comp_norm_inc}) and its equivalent expression for ray 4, together with eqs.~(\ref{eq: x_comp_init_cond_ray_1}) and (\ref{eq: x_comp_init_cond_ray_3}), we are able to write (making $c\neq 1$) the interference pattern as a function of time:
\begin{subequations} \label{eq:finalintensity}
\begin{align}
&I(t_{\mathcal{D}}) =  (\omega_{\textrm{e}\mathcal{E}} \mathcal{E}^x)^2 \bigg[\frac{\cos{[\omega_{\textrm{e}\mathcal{E}}(t_{\mathcal{D}} - 2\Delta \ell'/c)]}}{c\widehat{\Theta}_{\mathcal{E}} \Delta \ell' + \omega_{\textrm{e}\mathcal{E}}}  - \frac{\cos{[\omega_{\textrm{e}\mathcal{E}}(t_{\mathcal{D}} - 2\Delta \ell/c)]}}{c\widehat{\Theta}_{\mathcal{E}} \Delta \ell + \omega_{\textrm{e}\mathcal{E}}}\bigg] \times  \nonumber \\ & \hspace{67pt} \times \bigg \{ \frac{\cos{[\omega_{\textrm{e}\mathcal{E}}(t_{\mathcal{D}} - 2\Delta \ell'/c)]}}{c\widehat{\Theta}_{\mathcal{E}} \Delta \ell' + \omega_{e{\mathcal{E}}}} - \frac{\cos{[\omega_{\textrm{e}\mathcal{E}}(t_{\mathcal{D}} - 2\Delta \ell/c)]}}{c\widehat{\Theta}_{\mathcal{E}} \Delta \ell + \omega_{\textrm{e}\mathcal{E}}} + T'(t_{\mathcal{D}}) - T(t_{\mathcal{D}}) \bigg\}\,,
\end{align}
with:
\begin{align}
T(t_{\mathcal{D}}) \leqdef & \frac{4\omega_{\textrm{e}\mathcal{E}}}{c} \frac{\Delta L(ct_{\mathcal{D}} - \Delta \ell)}{c \widehat{\Theta}_{\mathcal{E}} \Delta \ell + \omega_{\textrm{e}\mathcal{E}}} \sin{[\omega_{\textrm{e}\mathcal{E}}(t_{\mathcal{D}} - 2 \Delta \ell/c)]} +  \frac{2 \cos{[\omega_{\textrm{e}\mathcal{E}}(t_{\mathcal{D}} - 2\Delta \ell/c)]}}{c\widehat{\Theta}_{\mathcal{E}}\Delta \ell + \omega_{\textrm{e}\mathcal{E}}} \frac{\Delta \omega_{\textrm{e}}}{\omega_{\textrm{e}\mathcal{E}}}(t_{\mathcal{D}}) \nonumber \\
& - \frac{2c \widehat{\Theta}_{\mathcal{E}} }{[c\hat{\Theta}(t_{\mathcal{E}}) \Delta \ell + \omega_{\textrm{e}\mathcal{E}}]^2} \cos{[\omega_{\textrm{e}\mathcal{E}}(t_{\mathcal{D}} - 2 \Delta \ell/c)]} \bigg\{2 \Delta L(ct_{\mathcal{D}} - \Delta \ell) \nonumber \\ & +  \frac{(\Delta y^2 - \Delta z^2)}{2\Delta \ell} \epsilon_+ h_{+}(ct_{\mathcal{D}} - 2\Delta \ell- x_{\mathcal{S}}) + \frac{\Delta y \Delta z}{\Delta \ell} \epsilon_{\times}  h_{\times}(ct_{\mathcal{D}} - 2\Delta \ell - x_{\mathcal{S}})    \bigg\}, \label{eq:non_trivial_pert}
\end{align}
\end{subequations}
where $T'$ has the same expression but for the other arm, namely, with the changes $\Delta \ell \rightarrow \Delta \ell'$, $D_R \rightarrow D_R'$, $\Delta y \rightarrow \Delta y'$, $\Delta z \rightarrow \Delta z'$. Here $\Delta\omega_{\textrm{e}}/\omega_{\textrm{e}\mathcal{E}}$ is the Doppler shift after a round-trip travel of light, explicitly calculated on L1 to be, for normal GW incidence
\begin{align}
\frac{\Delta \omega_{\textrm{e}}}{\omega_{\textrm{e}\mathcal{E}}} (t_{\mathcal{D}})= \, &  \epsilon_{\times}\frac{\Delta y \Delta z}{\Delta\ell^2} [ h_{\times}(ct_{{\mathcal{D}}} - x_{\mathcal{M}}) - h_{\times}(ct_{{\mathcal{D}}} - 2 \Delta \ell- x_{\mathcal{S}})]  + \nonumber \\ & \epsilon_+ \frac{\Delta y^2 - \Delta z^2}{2\Delta\ell^2}  [ h_{+}(ct_{{\mathcal{D}}} - x_{\mathcal{M}})- h_{+}(ct_{{\mathcal{D}}} - 2 \Delta \ell- x_{\mathcal{S}})]\,. \label{eq:freq_shift}
\end{align}

Eqs.~(\ref{eq:finalintensity}) are the main results of this second part of our work. They give the instantaneous intensity measured at the end of the interferometry process in our toy model detector. By making $\bm{\epsilon} = 0$, we notice that the only non-vanishing term is the difference in cosines squared. These are the Minkowski contributions and, when $\widehat{\Theta}_{\mathcal{E}} = 0$, can be combined to give the usual single sine squared of the difference of the arms' lengths \cite{Maggiore2007}. 

The contributions of $T$ and $T'$ arise from the interaction of GWs with the laser beams and are all of linear order on the parameter $\bm{\epsilon}$. The first term in eq.~($\ref{eq:non_trivial_pert}$) and the equivalent one for the other arm are the traditional perturbations obtained when discussing the detection of GWs and are associated with the phase difference of the interacting beams. In fact, they come from the phase of the initial conditions in eqs.~(\ref{eq: x_comp_init_cond_ray_1}) and ($\ref{eq: x_comp_init_cond_ray_3}$) and are characterized mainly by the anisotropic change in the arms' radar lengths ($\Delta L$ and $\Delta L'$) which results in a difference of optical paths along them. We see that new effects are also present. One is proportional to the frequency shift arising from the time variation of the radar distance between the arm's extremities (see L1 for a complete interpretation) and the others are proportional to the initial value of the optical expansion parameter of the beams. Both were expected to influence such final intensity by their explicit presence on eq.~(\ref{eq:intensity_evolution}), terms whose physical origins were previously brought to light (cf. section \ref{sec:electriceq}). It is important to emphasize, as discussed in L1, that the frequency shift does not contribute to any phase shift along the rays, since the phase on each ray is constant as the geometrical optics regime demands to be. Actually, the Doppler shift contribution to the interference pattern does not originate from the phase of the eletromagnetic wave, as is the case for the traditional term, but it is only a consequence of the electric field magnitude non-trivial propagation.

We note that although our experiment is set up in the dark fringe, the square bracket factor on (\ref{eq:finalintensity}) informs us that if the unperturbed arms have equal lengths, \emph{i.e. $\Delta \ell = \Delta \ell'$}, then $I(t_{\mathcal{D}}) = 0$, even if GWs are present (and thus $\Delta L \neq \Delta L'$). This is indeed what happens in inertial frames of Minkowski spacetime, that is, the interference pattern is quadratic in the difference of arms' lengths when it is small compared to the EM wavelength, and so a first conjecture is that, if this difference was only caused by the GW and yet the functional form of the intensity was the same $I(t_{\mathcal{D}}) \sim (\Delta L - \Delta L')^2$, there should be no contributions up to linear order in $\bm{\epsilon}$. Of course, as we derived, the form of the intensity changes itself by additional terms, but this property still holds in our toy model interferometer.

We seek now to compare the relevance of each of the contributions on eq.~(\ref{eq:non_trivial_pert}). We will compare the terms on the $T$ function but a completely analogous analysis may be carried out for the $T'$ contributions. Factoring out the common factor $2/(c\widehat{\Theta}_{\mathcal{E}}\Delta \ell + \omega_{e\mathcal{E}})$, we organize the amplitudes of the non-trivial contributions: 
\begin{align}
&C_1 \leqdef \frac{2\omega_{\textrm{e}\mathcal{E}}}{c} \Delta L(ct_{\mathcal{D}} - \Delta \ell), \end{align} \begin{align}
 &C_2 \leqdef  \frac{\Delta \omega_{\textrm{e}}}{\omega_{\textrm{e}\mathcal{E}}}(t_{\mathcal{D}}), \\ &	
C_3\leqdef - \frac{2c \widehat{\Theta}_{\mathcal{E}}\Delta L(ct_{\mathcal{D}}- \Delta \ell) }{c\widehat{\Theta}_{\mathcal{E}} \Delta \ell + \omega_{\textrm{e}\mathcal{E}}}, 
\\
&C_4 \leqdef \frac{c X_P  \widehat{\Theta}_{\mathcal{E}} \Delta \ell  }{c\widehat{\Theta}_{\mathcal{E}} \Delta \ell + \omega_{\textrm{e}\mathcal{E}}} \epsilon_P h_P(ct_{\mathcal{D}} - 2\Delta \ell -x_{\mathcal{S}}),   
\end{align}
with $X_P$ symbolizing the coordinate factors:
\begin{align}
X_+ \leqdef \frac{\Delta y^2 - \Delta z^2}{2\Delta \ell^2}\,, \quad X_{\times} \leqdef \frac{\Delta y \Delta z}{\Delta \ell^2}\,.
\end{align}

We begin by comparing  $C_1$ and $C_3$:
\begin{align}
	\frac{|C_1|}{|C_3|} &= \frac{\omega_{e \mathcal{E}}\Delta \ell}{c} \left( 1 + \frac{\omega_{\textrm{e}\mathcal{E}}}{c \widehat{\Theta}_{\mathcal{E}}\Delta \ell}\right) > \frac{\omega_{e \mathcal{E}}\Delta \ell}{c} = \frac{2 \pi \Delta \ell}{\lambda_{e \mathcal{E}}} \gg 1\,,
\end{align}
since we assume divergent beams, so that $\widehat{\Theta}_{\mathcal{E}}>0$. For aLIGO, $\mathcal{O}(C_1/C_3) = 10^{11}$, remembering eq.~(\ref{eq:init_Theta_est}) and that $\Delta \ell = 4$ km and $\omega_{e \mathcal{E}} = 1.8 \times 10^{15}$ rad/s \cite{Martynov2016}. Thus, $\mathcal{O}(C_1) \gg \mathcal{O}(C_3)$.

The remaining comparisons cannot be made in the same straightforward way.  Instead, we first assume our GW to be a general wave packet given by the Fourier decomposition:
\begin{equation}
h_P(ct-x) = \textrm{Re}\bigg\{\int_{- \infty}^{\infty} \tilde{h}_P(\omega_g) \textrm{e}^{i\frac{\omega_g}{c}(ct - x)}d\omega_g \bigg\}, \label{eq:wave_packet}
\end{equation}
where $\textrm{Re}(\alpha)$ is the real part of $\alpha$. Then, by eq.~(\ref{eq:freq_shift}), we know that a representative term of the frequency shift present in $C_2$ is of the form:
\begin{align}
	\frac{\Delta \omega_{\textrm{e}}}{\omega_{\textrm{e}\mathcal{E}}}(t_{\mathcal{D}}) & \sim \epsilon_P X_P [h_P(ct_{\mathcal{D}} -x_{\mathcal{S}}) - h_P(ct_{\mathcal{D}} - 2\Delta \ell -x_{\mathcal{S}})] \nonumber \\
	&= -2\epsilon_P X_P \int \sin\Big(\frac{\Delta \ell}{\lambdabar_g}\Big) \bigg[\textrm{Re}(\tilde{h}_P)\sin\bigg(\frac{ct_{\mathcal{D}} - \Delta \ell- x_{\mathcal{S}}}{\lambdabar_g}\bigg) \nonumber \\ & \hspace{107pt} +\textrm{Im}(\tilde{h}_P)\cos\bigg(\frac{ct_{\mathcal{D}} - \Delta \ell - x_{\mathcal{S}}}{\lambdabar_g}\bigg)\bigg]   d\omega_g\,,
	 \label{eq:freq_modes}
\end{align}
where $\textrm{Im}(\alpha)$ is the imaginary part of $\alpha$, $\lambdabar_g \leqdef \lambda_g/2 \pi$ is the reduced GW wavelength. As for the radar distance perturbation in $C_1$, one may attain the contributions of each mode in the same fashion by writing:
\begin{align}
	C_1  & = \frac{2 \omega_{\textrm{e}\mathcal{E}}}{c}\Delta L (ct_{\mathcal{D}} - \Delta \ell) \nonumber \\ & = 
	\frac{\epsilon_P \omega_{\textrm{e}\mathcal{E}} X_P }{c} \int_{ct_{\mathcal{D}} -2\Delta \ell -x_{\mathcal{S}}}^{ct_{\mathcal{D}}-x_{\mathcal{S}}}\textrm{Re}\left\{\int \tilde{h}_P(\omega_g) \textrm{e}^{i\frac{\omega_g}{c}w}d\omega_g \right\}dw
	\nonumber \\&  =2 \epsilon_P X_P \int \frac{\omega_{\textrm{e}\mathcal{E}}}{\omega_g} \sin\bigg(\frac{\Delta \ell}{\lambdabar_g}\bigg)  \bigg[\textrm{Re}(\tilde{h}_P) \cos\left(\frac{ct_{\mathcal{D}} - \Delta \ell - x_{\mathcal{S}}}{\lambdabar_g}\right)  \nonumber \\ & \hspace{124pt}-\textrm{Im}(\tilde{h}_P) \sin\left(\frac{ct_{\mathcal{D}} - \Delta \ell - x_{\mathcal{S}}}{\lambdabar_g}\right)  \bigg]d\omega_g\,. \label{eq:radar_dist_modes}
\end{align}

Factoring out the common parameters of $C_1$ and $C_2$, we may compare these two terms by looking at the integrands in eqs.~(\ref{eq:freq_modes}) and (\ref{eq:radar_dist_modes}), which, apart from combinations of (bounded) harmonic functions, differ by a factor $\omega_{\textrm{e}\mathcal{E}}/\omega_g$. Since our description of the interferometry process is only valid in the electromagnetic geometrical optics regime and $\omega_g$ gives the scale of the metric variations (cf. appendix \ref{app:geometrical_optics}), we must have, for each GW mode
\begin{align}
\frac{|C_1|}{|C_2|} \sim \frac{\omega_{\textrm{e}\mathcal{E}}}{\omega_g} \gg 1\,.
\end{align}
In particular, for the aLIGO detectable spectrum, $\omega_e/\omega_g \geq 10^{11}$.

Contribution $C_4$ may be rewritten as
\begin{align} &C_4 = \frac{cX_P\epsilon_P\widehat{\Theta}_{\mathcal{E}}\Delta \ell}{c\widehat{\Theta}_{\mathcal{E}}\Delta \ell + \omega_{\textrm{e}\mathcal{E}}}\textrm{Re}\left\{\int \tilde{h}_P \textrm{e}^{i\frac{\omega_g}{c}(ct_{\mathcal{D} - 2\Delta \ell - x_{\mathcal{S}}})}d\omega_g \right\} = \frac{X_P\epsilon_P}{1 + \omega_{\textrm{e}\mathcal{E}}/c\widehat{\Theta}_{\mathcal{E}}\Delta \ell} \times \nonumber \\ \nonumber \\ & \times \int \bigg[ \textrm{Re}(\tilde{h}_P) \cos \bigg(\frac{ct_{\mathcal{D}} - \Delta \ell - x_{\mathcal{S}}}{\lambdabar_g}\bigg) - \textrm{Im}(\tilde{h}_P) \sin\left(\frac{ct_{\mathcal{D}} - \Delta \ell - x_{\mathcal{S}}}{\lambdabar_g}\right) \bigg]\cos\left(\frac{\Delta \ell}{\lambdabar_g}\right) \nonumber \\ \nonumber \\
& \hspace{13pt}+ \bigg[ \textrm{Re}(\tilde{h}_P) \sin \bigg(\frac{ct_{\mathcal{D}} - \Delta \ell - x_{\mathcal{S}}}{\lambdabar_g}\bigg) + \textrm{Im}(\tilde{h}_P) \cos\left(\frac{ct_{\mathcal{D}} - \Delta \ell - x_{\mathcal{S}}}{\lambdabar_g}\right) \bigg]\sin\left(\frac{\Delta \ell}{\lambdabar_g}\right). \label{eq:Theta_modes}
\end{align}
Comparing this expression with eq.~(\ref{eq:radar_dist_modes}), we conclude that the two last terms of eq.~(\ref{eq:Theta_modes}) are always smaller than $C_1$, again because of the frequency ratio present in the latter and the fact that
\begin{equation}
	\frac{1}{1+\omega_{\textrm{e}\mathcal{E}}/c\widehat{\Theta}_{\mathcal{E}}\Delta \ell} < 1\,.
\end{equation}
As for the first two terms in $C_4$, one arrives at the same conclusion, unless $\Delta \ell \approx n \lambda_g, n \in \mathbb{N}$,  because these modes are suppressed in $C_1$, but the corresponding ones in $C_4$ are not. On the particular case of the long-wavelength limit, $\Delta \ell /\lambda_g \ll 1$, we find:
\begin{equation}
	\frac{\omega_{\textrm{e}\mathcal{E}}}{\omega_g}\sin{\left(\frac{\Delta \ell}{\lambdabar_g}\right)} \approx \frac{\omega_{\textrm{e}\mathcal{E}}}{\omega_g} \frac{\Delta \ell}{\lambdabar_g} = \frac{\Delta \ell}{\lambdabar_e} \gg 1\,,
\end{equation}
while
\begin{equation}
		\frac{1}{1+\omega_{\textrm{e}\mathcal{E}}/c\widehat{\Theta}_{\mathcal{E}}\Delta \ell} \cos{\left(\frac{\Delta \ell}{\lambdabar_g}\right)} \approx 	\frac{1}{1+\omega_{\textrm{e}\mathcal{E}}/c\widehat{\Theta}_{\mathcal{E}}\Delta \ell} < 1\,.
\end{equation}
In the general case, then, one may conclude:
\begin{equation}
	\frac{|C_1|}{|C_4|} \gg 1\,.
\end{equation}

It is important to stress that for the above comparisons, we only assumed that the geometrical optics limit for light is valid and that the arms of the interferometer are much bigger than the laser wavelength. No assumption was necessary on the particular values of the parameters in question, allowing us to conclude that the dominant term is always the traditional one, for the normal incidence of GWs. We remember that our model assumes passive reflection and so it should be modified for the LISA detector. We also emphasize that the comparisons made were between the instantaneous values of each contribution and a more accurate treatment could be achieved by taking their time average.

\begin{figure}[t] \label{fig:contributions}
	\centering
	\includegraphics[scale=1.0]{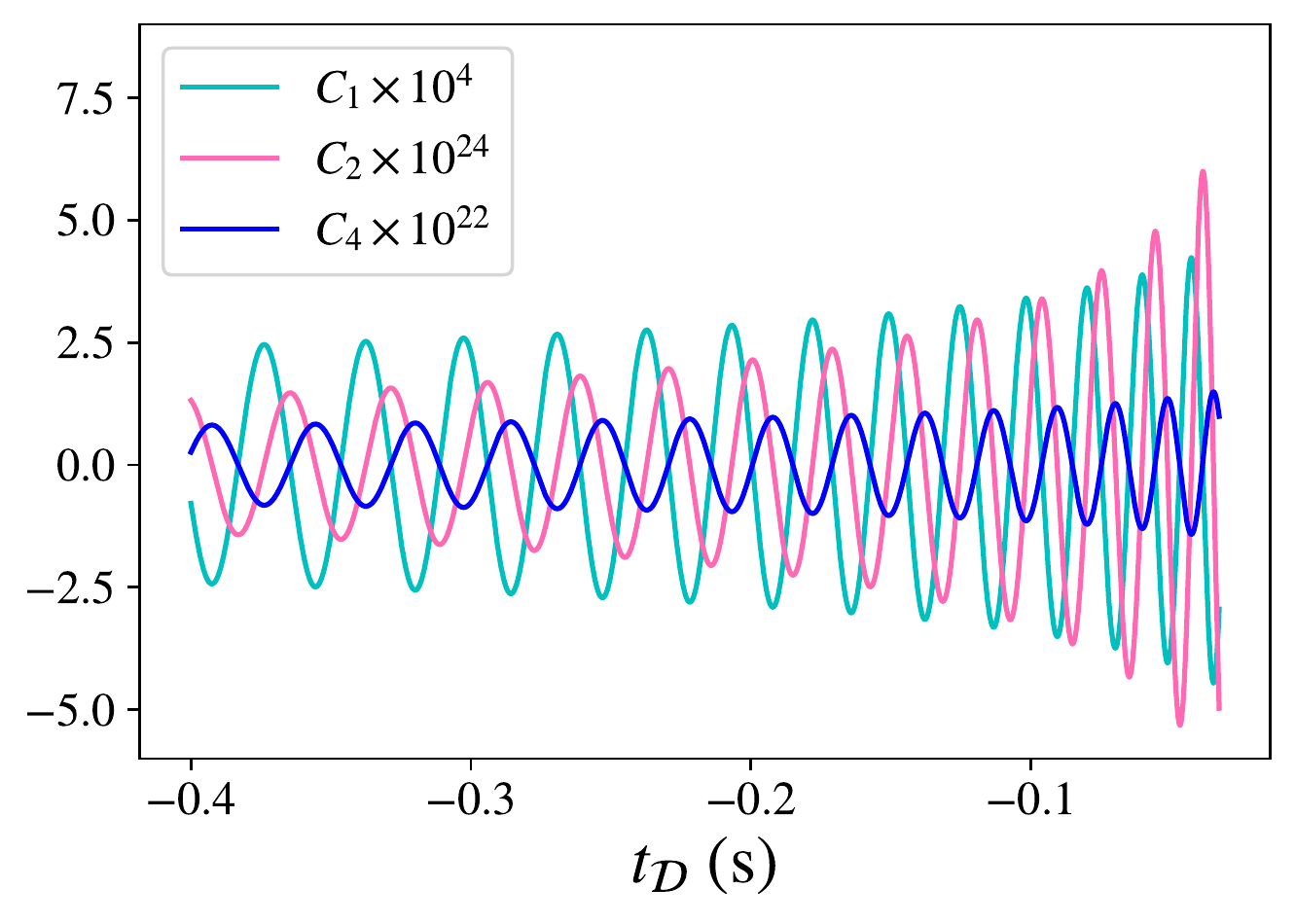}
	\caption{Contributions to the interference pattern as functions of the detection time for a typical GW signal observed by aLIGO corresponding to the inspiral phase of a binary source (up to $\omega_g \simeq 425$ rad/s). We have made $\theta = \pi/2$, $\phi = \pi/3$, $\Delta \ell =$ 4 km, $M_c = 28.3 \, \text{M}_{\odot}$, $r = 410$ Mpc and $\Phi_0=t_c=\iota = 0$.}
\end{figure}

Although the above comparisons were made in a much more general case, we present on figure \ref{fig:contributions} the amplitudes $C_1$, $C_2$ and $C_4$ for the same simple template of the GW amplitude used in L1 (see its figure 4). It corresponds to an approximated signal emitted by a binary source whose chirp mass $M_c$ and luminosity distance $r$ were chosen to be the ones inferred from the first detection of GWs by aLIGO, with all the other parameters set to vanish. $C_3$ was not shown since it is simply proportional to $C_1$. We note that, as anticipated, the non-traditional contributions can be safely neglected.

\section{Conclusion}

As a sequence of our previous work L1, in this paper we investigated a third facet of the effect of GWs in light, namely, their influence on the electric field propagation along light rays in interferometric GW detectors. We applied our newly found eq.~(\ref{eq:electric_evolution}), which takes into account fully both the curved nature of spacetime in the EM geometrical optics regime and the kinematics of a chosen reference frame, to an idealized Michelson-Morley-like interferometer in the TT frame with the presence of a non-monochromatic plane GW traveling in a flat background, assuming passive reflection at the end mirrors. Then, as our key result, we were able to compute, in eq.~(\ref{eq:finalintensity}), for a normally incident GW, the final instantaneous EM interference pattern as a function of the GW amplitude, initial EM frequency, the laser beam divergence at emission, unperturbed arms' lengths and orientations. We found two new contributions besides the known traditional term related to the difference in optical paths: one associated to the frequency shift acquired by light during its round-trips in the arms, as a consequence of the TT kinematics, and other due to the expansion of the light beam. Despite being of linear order in the GW perturbative parameter $\bm{\epsilon}$, they showed to be negligible compared to the traditional contribution as long as the EM wavelength is much smaller than the GW one and the arms' lengths, conditions commonly understood as prerequisites for the validity of the geometrical optics approximation \cite{Schneider1992}. 
On the aLIGO case, we have estimated the value of the initial expansion parameter for the light beam and concluded that all non-traditional corrections are of order $10^{-11}$ when compared with the traditional one.

A third new contribution to the interfered light intensity was foreseen to arise from the non-parallel transport of the EM polarization vector. For GW normal incidence we have shown that such vector is indeed parallel transported and, thus, this contribution is not present, but in a more general case, it is expected to perturb the measured signal. Moreover, ignoring such vector nature of the electric field in this case allows us to assess the interference pattern a priori, by only looking at the intensity evolution in the arms. For this, it is enough to consider, in view of eq.~(\ref{eq: final_intensity_simplification}), that the electric fields before superposition at detection are given by
\begin{align}
E(t_{\mathcal{D}}) = \sqrt{I_{\mathcal{D}}}\,, \; E'(t_{\mathcal{D}}) = -\sqrt{I'_{\mathcal{D}}}\,,
\end{align}
and only use the intuitive eq.~(\ref{eq:brightness_conservation}) to write the final intensity at one arm
\begin{align}
I_{\mathcal{D}} = I_{\mathcal{E}}\frac{\delta S_{\mathcal{E}}}{\delta S_{\mathcal{D}}}\left(\frac{ \omega_{\textrm{e}\mathcal{D}}}{\omega_{\textrm{e}\mathcal{E}}}\right)^2\,,
\end{align}
where its initial value is a function of the phase times a constant amplitude, i.e. $I_{\mathcal{E}} = \mathcal{E}^2\cos(\psi_{\mathcal{E}})$. So $I_T = [E(t_{\mathcal{D}}) + E'(t_{\mathcal{D}})]^2$ becomes, in this simplifying reasoning,
\begin{align}
I_T = \mathcal{E}^2\Bigg[ \sqrt{\frac{\delta S_{\mathcal{E}}}{\delta S_{\mathcal{D}}}}\left(1 + \frac{\Delta \omega_{\textrm{e}}}{\omega_{\textrm{e}\mathcal{E}}}\right)\cos(\psi_{\mathcal{E}}) - \sqrt{\frac{\delta S_{\mathcal{E}}}{\delta S_{\mathcal{D}}'}}\left(1 + \frac{\Delta \omega_{\textrm{e}}'}{\omega_{\textrm{e}\mathcal{E}}}\right)\cos(\psi_{\mathcal{E}'})\Bigg]^2\,,
\end{align} 
which in fact agrees with eq.~(\ref{eq:finalintensity}), if one expands it up to linear order in $\bm{\epsilon}$ and relate, by eq.~(\ref{Thetaandarea}), the ratio of areas with $\widehat{\Theta}$. In this way we definitely see that the newly obtained contributions come from the intensity evolution instead of the phase, as discussed previously (cf. section \ref{sec:finalintensity} and L1).

Under the assumptions of this work, then, although several features regarding perturbations in light induced by GWs do ensue in linear order, e.g. spatial trajectory deviations, Doppler effect, polarization tilts and intensity fluctuations, one may certify that, at least for normal incidence, the detection of GWs justifiably relies on the interference pattern depending solely on the phase difference of the recombining rays, as presupposed in subsection V C of L1 and throughout most of literature.

\acknowledgments
JCL thanks Brazilian funding agencies CAPES and FAPERJ for MSc scholarships 31001017002-M0 and 2016.00763-4, respectively, and ISM thanks Brazilian funding agency CNPq for PhD scholarship GD 140324/2018-6.

\appendix
\section{Christoffel symbols}
\label{app:christoffel_symbols}
The Christoffel symbols of the metric (\ref{eq:metric}) can be simplified, when dealing with approximations up to linear order in $\bm{\epsilon}$, to:
\begin{align}
\Gamma^{t}_{\beta \gamma} &= \frac{\epsilon_P}{2} h^P_{\beta \gamma,t}\,, \label{gamma^t}\\
\Gamma^{i}_{t j} &= \frac{\epsilon_P}{2} h^P_{j i,t}\,, \label{gamma_t}\\
\Gamma^{i}_{x j} &= \frac{\epsilon_P}{2} h^P_{j i,x}\,, \label{gamma_x}\\
\Gamma^{i}_{y j} &= \epsilon_P h^P_{y [i,j]}\,, \label{gamma_y}\\
\Gamma^{i}_{z j} &= \epsilon_P h^P_{z [i,j]}. \label{gamma_z}
\end{align}
\section{Geometrical optics approximation of Maxwell equations}
\label{app:geometrical_optics}

The geometrical optics approximation of Maxwell's equations in vacuum,
\begin{subequations}
	\label{eq:maxwell}
	\begin{eqnarray}
	{F^{\mu\nu}}_{;\nu} &=& 0\,, \label{eq:gauss} \\
	F_{[\mu\nu;\lambda]} &=& 0\,,  \label{eq:faraday}
	\end{eqnarray}
\end{subequations}
is established by searching for solutions of these field equations in the form of a one-parameter ($\eta$) 
family of electromagnetic fields \cite{Ehlers1967, Misner1973, Schneider1992, Perlick2000, Ellis2012, Harte2019}:
\begin{subequations}
	\label{eq:ansatz}
	\begin{eqnarray}
	F_{\mu \nu}(x, \eta) &=& f_{\mu \nu}(x, \eta) \textrm{e}^{i \psi(x)/\eta}\,, \label{eq:ansatz_product} \\
	f_{\mu \nu}(x, \eta) &\leqdef& \sum_{n = 0}^N f_{(n)\mu \nu}(x) \eta^n \quad (N \ge 0)\,. 	\label{eq:amplitude}
	\end{eqnarray}
\end{subequations}

\begin{table*}[t]
	\centering
	\begin{tabular}{ccc}
		Order & Final equation from (\ref{eq:gauss}) & Final equation from (\ref{eq:faraday})\\
		\hline \\
		$\eta^{-1}:$ & ${f_{(0)}}^{\mu\nu}\tilde{k}_\nu = 0$ &  $f_{(0)[\mu\nu}\tilde{k}_{\lambda]}=0$ \\
		\\
		\hline \\
		$\eta^p\quad (0\leq p\leq N-1):$ &  ${{f_{(p)}}^{\mu\nu}}_{;\nu} + i{f_{(p+1)}}^{\mu\nu}\tilde{k}_\nu = 0$   & $f_{(p)[\mu\nu;\lambda]} + if_{(p+1)[\mu\nu}\tilde{k}_{\lambda]} = 0$ \\
		\\
		\hline \\
		$\eta^N$: &  ${{f_{(N)}}^{\mu\nu}}_{;\nu} = 0$   & $f_{(N)[\mu\nu;\lambda]} = 0$
	\end{tabular}
	\caption{\label{tab:hierarchy}Hierarchy of Maxwell's equations for the geometrical optics approximation.}		
\end{table*}

{In general, $f_{\mu\nu}$ is a complex antisymmetric smooth tensor field 
and $\psi(x)/\eta$ is a real smooth scalar field; these are called, respectively, the amplitude 
and phase of the electromagnetic wave; $\eta$ is a dimensionless perturbation parameter proportional to the 
wavelength of the electromagnetic wave. Naturally, the real part of $F_{\mu \nu}$ must be taken in the end. This \emph{Ansatz} 
generalizes the plane wave monochromatic solution of Maxwell's equations in Minkowski spacetime (in pseudo-Cartesian 
coordinates, adapted to  an inertial frame of reference), and is expected to represent, in the limit $\eta 
\to 0$, a rapidly oscillating function of its phase, with a slowly varying amplitude. Moreover, the vector field 
defined by
\begin{equation} 
\label{wave-vector}
\tilde{k}_{\mu}(x) \leqdef  \psi_{,\mu}(x)
\end{equation} 
is supposed to have no zeros in the considered region (irrespective of the values of $\eta$), and 
\begin{equation}
\label{physical_wavevector}
k_\mu \leqdef \frac{\tilde{k}_\mu}{\eta}
\end{equation}
should be interpreted as the wave vector field of the electromagnetic wave, proportional to the momentum field of a 
stream of photons. Finally, $f^{(0)}_{\mu \nu}$ is assumed to vanish  at most in a set of measure zero. 
Inserting eq.~(\ref{eq:ansatz}) into Maxwell's equations (\ref{eq:maxwell}) and demanding their validity for all 
values of $\eta$, we find $N + 1$ hierarchical relations, the first two of them given by (cf. Table 
\ref{tab:hierarchy}):
\begin{itemize}
	\item dominant $\eta^{-1}$ order:
\end{itemize}
\begin{subequations}
	\label{eta-1}
	\begin{eqnarray}
	{f_{(0)}}^{\mu \nu} \tilde{k}_\nu   & = & 0\,,\label{eq:gauss-1} \\
	f_{(0)[\mu \nu} \tilde{k}_{\lambda]} & = & 0\,, \label{eq:faraday-1}
	\end{eqnarray}
\end{subequations}
and
\begin{itemize}
	\item subdominant $\eta^0$ order:
\end{itemize}
\begin{subequations}
	\label{eq:epsilon0}
	\begin{eqnarray}
	{{f_{(0)}}^{ \mu \nu}}_{;\nu} + i \tilde{k}_\nu {f_{(1)}}^{\mu \nu} & = & 0\,, \label{eq:gauss0} \\
	f_{(0)[\mu \nu;\lambda]} + i  f_{(1)[\mu \nu}\tilde{k}_{\lambda]} & = & 0\,. \label{eq:faraday0}
	\end{eqnarray}
\end{subequations}

Projecting eq.~(\ref{eq:faraday-1}) onto $\tilde{k}^{\mu}$ and taking eq.~(\ref{eq:gauss-1}) into account, it 
immediately follows that
\begin{equation}
\label{eq:null_tangent}
\tilde{k}_{\mu} \tilde{k}^{\mu} = 0 = k_\mu k^\mu\,,
\end{equation}
which implies that the integral curves of the wave vector field (or rays) are null curves, and, 
together with eq.~(\ref{wave-vector}), that the surfaces of constant phase are null hypersurfaces. 
Besides, since $\tilde{k}^{\mu}$ has vanishing curl, these equations also show that the rays are geodesics, 
also demonstrating that the light rays form a bundle with zero optical vorticity. 
These are consistent with the results one obtains when studying the characteristic surfaces and 
bi-characteristic curves of Maxwell's equations in vacuum \cite{Lichnerowicz1960}, or even when considering 
shock waves of the electromagnetic field \cite{Papapetrou1977}.

Besides, projecting eq.~(\ref{eq:faraday0}) onto $\tilde{k}^{\mu}$, and using eqs.~(\ref{eq:gauss-1}), (\ref{eq:faraday-1}), 
(\ref{eq:gauss0}) and (\ref{eq:null_tangent}), we get
\begin{equation}
\label{eq:transport_zero_order}
f_{(0)\mu \nu;\lambda}\tilde{k}^{\lambda} + \frac{1}{2} {\tilde{k}^{\lambda}}_{;\lambda}  f_{(0)\mu \nu} = 0\,.
\end{equation}
Here, we note that the above equation gives the evolution of $f^{(0)}_{\mu \nu}$ independently of 
the other $f^{(p)}_{\mu \nu}$ ($p = 1, ..., N$). The usual geometrical optics approximation relies on taking $ \eta \to 0 $, and assuming 
that $f^{(0)}_{\mu \nu}$ is a good approximation for the amplitude of the electromagnetic field, in which case all other
contributions may be disregarded. This assumption 
translates the physical demand that $F_{\mu\nu}(x,\eta)$ is to vanish at an arbitrarily large 
discrete number of hypersurfaces (``nodes''), so that it can be interpreted as a realistic wave.
If we stick to it, $F_{\mu \nu} (x, \eta) \approx f_{(0)\mu \nu}(x) \textrm{e}^{i \psi(x)/\eta}$, and the
higher-order corrections are to be neglected \cite{Harte2019}}. Then, eqs.~(\ref{eta-1}) and (\ref{eq:transport_zero_order}) become, 
respectively, equivalent to
\begin{subequations}
	\begin{eqnarray}
	k_\nu F^{\mu \nu} & = 0\,, \label{faraday-eigen} \\
	k_{[\lambda} F_{\mu \nu]} & = 0\,, \label{hodge-eigen}
	\end{eqnarray}
\end{subequations}
and
\begin{equation}
\label{faraday-transport-equation}
F_{\mu \nu;\lambda}k^{\lambda} + \frac{1}{2} {k^{\lambda}}_{;\lambda} F_{\mu \nu} = 0\,,
\end{equation}
where we have already included $\eta$ in the previous two equations, since they both contain the same orders of $\tilde{k}_\mu$.
eqs.~(\ref{faraday-eigen}) and (\ref{hodge-eigen}) show that the wave vector is a principal 
null direction of both the electromagnetic field and its dual, and the considered \emph{Ansatz} 
corresponds (approximately) to a null electromagnetic field \cite{Hall2004, Harte2019}. 
Last, eq.~(\ref{faraday-transport-equation}) shows how the electromagnetic field is transported 
along any of its associated light rays, and, together with eq.~(\ref{hodge-eigen}), is the path 
leading to the transport equation for the electric field appearing in \cite{Santana2020} and to the results presented here therefrom.

Using the above constraints, we can establish a condition for the validity of the geometrical optics regime in the particular spacetime we use throughout this series, namely, a GW perturbed Minkowski background. For this, we replace $F_{\mu \nu} (x, \eta) \approx f_{(0)\mu \nu}(x) \textrm{e}^{i \psi(x)/\eta}$ in eq.~(\ref{eq:gauss}) and find:
\begin{equation}
\frac{i\tilde{k}_{\nu}f_{(0)}^{\mu \nu}}{\eta} + f_{(0),\nu}^{\mu \nu}  + \Gamma^{\nu}_{\alpha \nu} f_{(0)}^{\mu \alpha}= 0\,. \label{eq:go_app}
\end{equation}
In the particular case of our GW spacetime, in addition to $1/\eta$, there are two other expansion parameters $(\epsilon_+, \epsilon_{\times})$ that need to be taken into account simultaneously. In order to obtain the hierarchical relations (\ref{eq:gauss-1}), we need to check whether the last two terms in the above equation can be neglected as compared to the first one. Indeed, one of the geometrical optics assumptions is that the amplitude of the Faraday tensor varies much less than its phase and thus the second term (both imaginary and real parts) is considered to be much smaller than the first one. Furthermore, remembering that
\begin{align}
\tilde{k}_{\nu} = \tilde{k}_{\nu (\bm{0})} + \epsilon_P\tilde{k}^P_{\nu} = \eta k_{\nu}\,,
\end{align}
eq.~(\ref{eq:go_app}) can be written as having contributions of three orders, namely
\begin{align}
\left[\left(\frac{1}{\eta}\right)i\tilde{k}_{\nu(\bm{0})} + \left(\frac{\epsilon_P}{\eta}\right)i\tilde{k}_{\nu}^P + \Gamma^{\alpha}_{\nu \alpha}\right]f_{(0)}^{\mu \nu} \approx 0\,, \label{eq:geo_opt_app}
\end{align}
where the Christoffel symbols are of order $\epsilon_P\omega_g$ for each mode in eq.~(\ref{eq:wave_packet}). Since $\mathcal{O}(\tilde{k}^P/\eta) = \omega_{\textrm{e}\mathcal{E}}$, the crossed term is of order $\epsilon_P\omega_{\textrm{e}\mathcal{E}}$ and, therefore, we must have
\begin{align}
\frac{\omega_{\textrm{e}\mathcal{E}}}{\omega_g} \gg 1\,, \label{eq:freq_ratio}
\end{align}
so that the last term is negligible compared to the others and thus the transversality of the EM field under the geometrical optics regime and all of its consequences are guaranteed. Of course a more rigorous and similar argument can be made by comparing the order of magnitude of the contributions in the real and imaginary parts of eq.~(\ref{eq:geo_opt_app}), which would ultimately lead to the same conclusions. Condition (\ref{eq:freq_ratio}) is in agreement with the usual statement that the validity of the geometrical optics limit resides in EM wavelengths much smaller than the other relevant lengths of the system in question.

\bibliographystyle{unsrt}
\bibliography{interaction_EM_waves_and_GW_part_2}
\end{document}